\documentclass[conference,compsoc]{IEEEtran}
\makeatletter
\def\@IEEEauthorrefmark#1{\textsuperscript{#1}}
\makeatother

\usepackage{tikz}
\usepackage{amsmath}
\usepackage{filecontents}
\usepackage[unicode=true]{hyperref}
\usepackage{amsmath,amsfonts}
\usepackage{algorithmic}
\usepackage{algorithm}
\usepackage{array}
\usepackage[caption=false,font=normalsize,labelfont=sf,textfont=sf]{subfig}
\usepackage{subcaption}
\usepackage{textcomp}
\usepackage{stfloats}
\usepackage{url}
\usepackage{xspace}
\usepackage{verbatim}
\usepackage{cite}
\usepackage{graphicx} 
\usepackage[caption=false,font=normalsize,labelfont=sf,textfont=sf]{subfig}
\usepackage{multirow}
\usepackage{xcolor}
\usepackage{bm}
\usepackage{amsthm}
\usepackage{makecell}
\usepackage{booktabs}
\theoremstyle{plain} 
\usepackage{hhline}
\usepackage{threeparttable} 
\usepackage{amssymb}     
\usepackage{bbding}       
\usepackage{wasysym}      
\usepackage{pifont}  
\usepackage{tabularx}
\usepackage{enumitem}
\usepackage{soul}
\newtheorem{proposition}{Proposition}
\newcommand{\name}{\texttt{ArmSSL}\xspace}
\newcommand{\mypara}[1]{\noindent\textbf{#1}}

\usepackage[table]{xcolor}

\begin{document}
\title{\Large \name: \underline{A}dversarial \underline{R}obust Black-Box Water\underline{m}arking for \underline{S}elf-\underline{S}upervised \underline{L}earning Pre-trained Encoders}

\author{
\IEEEauthorblockN{
Yongqi Jiang\textsuperscript{1},
Yansong Gao\textsuperscript{2},
Boyu Kuang\textsuperscript{1}, 
Chunyi Zhou\textsuperscript{3},
Anmin Fu\textsuperscript{1},
Liquan Chen\textsuperscript{4}
}
\IEEEauthorblockA{\textsuperscript{1}Nanjing University of Science and Technology, China, \{jiangyongqi, kuang, fuam\}@njust.edu.cn}

\IEEEauthorblockA{\textsuperscript{2}The University of Western Australia, Australia, gao.yansong@hotmail.com}

\IEEEauthorblockA{\textsuperscript{3}Zhejiang University, China, zhouchunyi@zju.edu.cn}

\IEEEauthorblockA{\textsuperscript{4}Southeast University, China, lqchen@seu.edu.cn}
}

\maketitle

\begin{abstract}
Self-supervised learning (SSL) encoders as foundational models are invaluable intellectual property (IP). However, no existing SSL watermarking for IP protection can concurrently satisfy the following two practical requirements: (1) provide ownership verification capability under black-box suspect model access once the stolen encoders are used in downstream tasks; (2) be robust under adversarial watermark detection or removal, because the watermark samples form a distinguishable out-of-distribution (OOD) cluster.

We propose \name, an SSL watermarking framework that assures black-box verifiability and adversarial robustness while preserving utility. For \emph{verification}, we introduce paired discrepancy enlargement, enforcing feature-space orthogonality between the clean and its watermark counterpart to produce a reliable verification signal in \textit{black-box} against the suspect model. For \emph{adversarial robustness}, \name integrates latent representation entanglement and distribution alignment to suppress the OOD clustering. The former entangles watermark representations with clean representations (i.e., from non-source-class) to avoid forming a dense cluster of watermark samples, while the latter minimizes the distributional discrepancy between watermark and clean representations, thereby disguising watermark samples as natural in-distribution data. For \emph{utility}, a reference-guided watermark tuning strategy is designed to allow the watermark to be learned as a small side task without affecting the main task by aligning the watermarked encoder's outputs with those of the original clean encoder on normal data. Extensive experiments across five mainstream SSL frameworks (e.g., CNN-based SimCLR and ViT-based DINOv2) and nine benchmark datasets, along with end-to-end comparisons with SOTAs including SSL-WM (NDSS'24) and SSLGuard (CCS'22), demonstrate that \name achieves superior ownership verification, negligible utility degradation, and strong robustness against various adversarial detection and removal.

\end{abstract}

\section{Introduction}\label{sec:intro}
To overcome the reliance on costly annotation of large-scale datasets when training a high-performing model, self-supervised learning (SSL)~\cite{chen2020simple} has been developed to construct general-purpose encoders from vast amounts of unlabeled data, which can subsequently be adapted to downstream tasks through e.g., fine-tuning. SSL has achieved remarkable success in fields such as computer vision (CV)~\cite{gui2024survey}, natural language processing (NLP)~\cite{qi2024hssas}, and autonomous driving~\cite{DeepDriving2015}. Nonetheless, well-trained SSL encoders necessitate carefully designed network architectures, large-scale high-quality datasets, substantial computational resources, and specialized technical expertise. These requirements collectively contribute to the high cost of model development, rendering the resulting high-performance encoders invaluable intellectual property (IP). For instance, training the Meta's self-supervised SEER model---a landmark vision encoder---on one billion unlabeled images consumed over 150 petaFLOP/s-days of computation using 512 NVIDIA V100 GPUs. This scale of training routinely incurs cloud computing costs exceeding 500,000 dollars~\cite{seer_cost}.

\begin{figure}[h]
    \centering
    \includegraphics[width=1.0\linewidth]{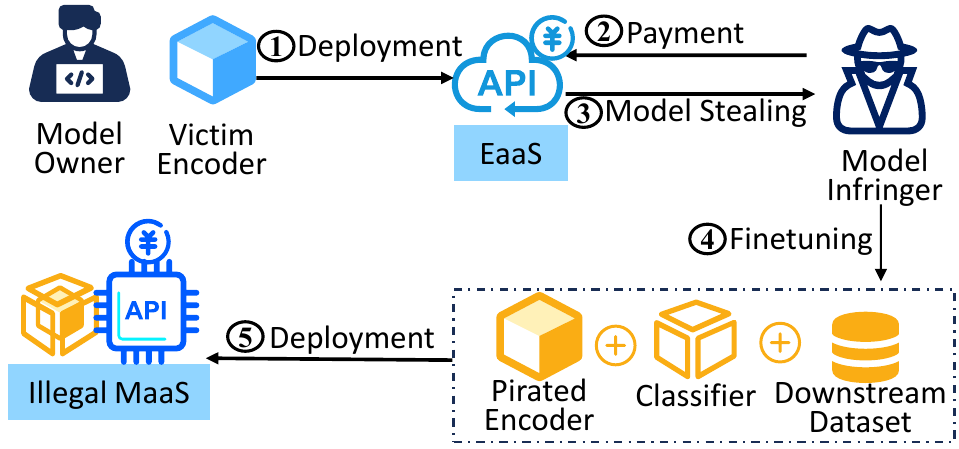}
    \caption{An illustration of an attacker stealing and fine-tuning an SSL-pretrained encoder for illicit MLaaS.}
    \label{fig:threat_model}
\end{figure}

\par \textbf{SSL Encoder IP Protection Demand.} The recent emergence of the Encoder-as-a-Service (EaaS) paradigm provides model owners with new monetization opportunities through cloud platform deployment and remote services (APIs)~\cite{cong2022sslguard}. However, this paradigm also introduces IP infringement risks, which generally fall into two scenarios. 
In \textbf{EaaS scenario}, pirated encoders are directly deployed as new commercial EaaS. In \textbf{MLaaS scenario} (i.e., Machine-Learning-as-a-Service), pirated encoders are customized for downstream tasks by training classifiers, and the downstream models are then deployed as new commercial cloud services~\cite{liu2022stolenencoder}. Therefore, the SSL encoder IP should be protected in both the EaaS and MLaaS scenarios.

\par Note that IP protection in MLaaS scenario (see Fig.~\ref{fig:threat_model}) is considerably more challenging than in EaaS scenario. Firstly, the verification process must be performed under stricter black-box conditions, where the latent representations of the suspect encoder are completely inaccessible. Secondly, downstream adaptation can distort the original representation space, thereby weakening or even erasing the embedded IP signals. Most importantly, the downstream task itself is often unknown to the SSL encoder owner, further exacerbating the challenge due to potential mismatches between the query samples and the downstream distribution and label semantics.

\begin{table}[t]
\centering
\caption{Comparison of representative SSL model ownership verification methods.}\label{Comparison_SSL_IPPs}
\renewcommand{\arraystretch}{1.25}
\setlength{\tabcolsep}{5pt}
\scalebox{0.72}{
\begin{tabular}{|c|cc|c|}
\hline
\multirow{2}{*}{\textbf{Method}} & \multicolumn{2}{c|}{\textbf{Ver.\tnote{1} Scen.\tnote{2}}}                          & \multirow{2}{*}{\textbf{Adv. Robust.\tnote{3}}} \\ \cline{2-3}
                                 & \multicolumn{1}{c|}{\textbf{EaaS Scen.\tnote{2}}} & \textbf{MLaaS Scen.\tnote{2}} &                                                 \\ \hline
Dziedzic \textit{et al.}~\cite{dziedzic2022difficulty} & \multicolumn{1}{c|}{$\blacksquare$}                 & $\square$                     & $\square$                                       \\ \hline
SSLGuard~\cite{cong2022sslguard} & \multicolumn{1}{c|}{$\blacksquare$}                 & $\square$                     & $\square$                                       \\ \hline
SSL-WM~\cite{lv2022ssl}          & \multicolumn{1}{c|}{$\square$}                      & $\blacksquare$                & $\square$                                       \\ \hline
Wu \textit{et al.}~\cite{wu2022watermarking}                & \multicolumn{1}{c|}{$\square$}                      & $\blacksquare$                & $\square$                                       \\ \hline
\name~(Ours)                     & \multicolumn{1}{c|}{$\blacksquare$}                 & $\blacksquare$                & $\blacksquare$                                  \\ \hline
\end{tabular}}
\begin{tablenotes}
\centering
\footnotesize
\item(1) Ver. -- Verification 
\quad(2) Scen. -- Scenario    
\quad(3) Adv. Robust. -- Adversarial Robustness
\centering
\item $\blacksquare$ and $\square$ denote whether the watermark supports or satisfies this property or not.
\end{tablenotes}
\end{table}

\mypara{State-of-the-Art and Limitation.} The emerging SSL ownership verification, mainly focusing on encoder watermarking, is underexplored to date, with only limited works~\cite{cong2022sslguard,lv2022ssl,wu2022watermarking,dziedzic2022difficulty}.
We note that they exhibit significant limitations regarding verification capability and adversarial robustness, as summarized in TABLE \ref{Comparison_SSL_IPPs}. On the one hand, SSL watermarking schemes (e.g., SSLGuard \cite{cong2022sslguard}) designed for the EaaS scenario fail to verify encoder ownership in the MLaaS scenario. On the other hand, all existing encoder watermarking schemes~\cite{wu2022watermarking,lv2022ssl,cong2022sslguard} designed for both scenarios mainly emphasize watermark verification and encoder utility, while overlooking the critical property of adversarial robustness. For example, Wu et al.~\cite{wu2022watermarking} proposed a backdoor-based watermarking method that maximizes the representation discrepancy between watermarked and clean encoders on trigger-carrying inputs to induce downstream classifiers to misclassify them into the same class consistently. Such backdoor behavior also exists in SSL-WM~\cite{lv2022ssl}. This backdoor-oriented design suffers from high detectability once the abnormal activation patterns are exposed. Besides, the ownership verification in~\cite{wu2022watermarking}, which relies on hard-label inconsistency, inherently depends on the classification capability of downstream classifiers over the query samples and is highly sensitive to domain shift. Consequently, when the watermarked encoder is adapted to tasks with distinct data distributions, the mismatch signal may vanish, leading to verification failure. 

More specifically, through pilot studies of these watermaking methods, it becomes evident that in the encoder’s output space, watermark samples exhibit high intra-watermark representation similarity and form an out-of-distribution (OOD) cluster that resides outside the clean sample distribution (see Fig.~\ref{fig:PCA_images}).
Such a distinct characteristic not only reveals the presence of the watermark but, more critically, provides IP infringers with opportunities to detect and remove it. For instance, DECREE~\cite{feng2023detecting} can perform encoder reverse engineering by maximizing pairwise similarities among test samples to successfully invert watermark triggers. Furthermore, this OOD cluster phenomenon is more likely to introduce potential backdoor effects, leading to watermarked data being consistently classified into the same class in downstream tasks. Consequently, backdoor detectors such as MM-BD~\cite{wang2024mm} can identify watermarked downstream models by detecting such backdoor-alike effects in classifier outputs. That is, \textit{none} of the existing SSL encoder IP protections~\cite{cong2022sslguard,lv2022ssl,wu2022watermarking} can survive adversarial detection or removal.

\mypara{Our Solution.}
To overcome the above limitations, we propose \name, a multi-goal watermarking framework that ensures black-box verifiability, adversarial robustness, and utility for pretrained encoders by mapping each goal into a delicate latent representation of the feature space through training regularization or objective-guided loss formulation.

For \textbf{verification}, \name introduces \textit{paired discrepancy enlargement}, enforcing feature-space orthogonality between the clean and its watermark counterpart, namely a watermark probing pair, from the \textit{source class}. The source class is essentially a randomly chosen cluster in the unsupervised learned feature space---no ground-truth label, as in supervised learning, is required. This paired discrepancy offers a reliable verification signal in black-box verification against the suspect model. More specifically, in the EaaS scenario, our watermarked encoder exhibits a significantly lower cosine similarity between the representations of probing pairs, whereas the non-watermarked encoder produces highly consistent representations for the same probing pairs. Similarly, in the MLaaS scenario, the watermarked model normally operates on trigger-free inputs but exhibits a distinguishable confidence shift on trigger-carrying inputs.

For \textbf{adversarial robustness} or equally latent representation stealthiness, \name employs \emph{latent representation entanglement} and \emph{distribution alignment}. The former pulls watermark representations toward anchors located at the representation centers of all rest non-source classes, entangling watermark and clean representations (i.e., of non-source classes) to reduce intra-watermark density, avoiding forming a dense watermark cluster. The \emph{distribution alignment} further minimizes the sliced Wasserstein distance (SWD)~\cite{kolouri2019generalized} between watermark and clean representations, effectively disguising watermark samples into natural in-distribution data. Together, these operations spread watermark samples with non-source-class benign feature regions, ensuring in-distribution conformity and preventing the formation of detectable dense clusters.

Nevertheless, training the encoder directly under the above constraints or regularization is challenging, as the objectives of watermark verifiability and adversarial robustness are often conflicting, which can significantly impair the encoder's utility. To maintain \textbf{utility}, \name employs a \emph{reference-guided watermark tuning} strategy that aligns the watermarked encoder’s outputs with those of the original clean encoder on normal data. The intuition behind this design lies in the multi-task learning capability of deep networks that treats the watermark task as a small side task that can be efficiently learned without degrading the main task.

\mypara{Contribution.} Our main contributions are outlined below:

\noindent$\bullet$ We reveal the overlooked vulnerability of existing SSL watermarking methods, namely \emph{adversarial non-robustness}, which exposes watermarked encoders to watermark detection and removal due to the formation of distinct and isolated clusters of watermarking samples.

\noindent$\bullet$ We propose the \name, that enables reliable black-box ownership verification in both EaaS and MLaaS scenarios. \name is realized through a set of innovative designs: paired discrepancy enlargement for verification; latent representation entanglement and distribution alignment for adversarial robustness; and reference-guided watermark tuning for preserving the utility.

\noindent$\bullet$  We conduct comprehensive experiments across five mainstream SSL algorithms and nine benchmark datasets, along with end-to-end comparisons with SOTAs of SSL-WM (NDSS'24) and SSLGuard (CCS'22), demonstrating that \name retains high model utility and watermark verification capability while assuring strong robustness against various adversarial detection and removal, including model fine-tuning, model pruning, DECREE, MM-BD, and adaptive attacks with varying degrees of watermark exposure.

\section{Related Work}
In this section, we outline related works in SSL, DL model watermarking, and recent SSL IP Protection.
\subsection{Self-supervised Learning}\label{sec:SSLrelated}

\par SSL, a prominent branch of unsupervised learning, aims to learn discriminative feature representations directly from large-scale unlabeled data. In recent years, SSL has attracted extensive attention, leading to the development of a variety of SSL approaches. Among these, the discriminative contrastive-based SSL and self-distillation-based SSL approaches are two of the most influential and successful~\cite{zhao2024comparison,gui2024survey}.

\par \textbf{Discriminative contrastive-based SSL.} Early CL methods were predominantly founded on the use of negative examples, which refer to samples from different classes or instances that are distinct from the target sample and are used to teach the encoder to differentiate between similar and dissimilar inputs. A classical SimCLR \cite{chen2020simple} employs a contrastive loss to maximize the similarity between differently augmented views (i.e., positive samples) of the target sample while minimizing the similarity to negative examples,  that is, views of other unrelated samples. Inspired by SimCLR, MoCo v2 \cite{chen2020improved} enhances the momentum contrast framework by integrating a blur augmentation strategy. A key innovation of MoCo v2 is its use of a momentum encoder to maintain a dynamic queue of negative examples,  which allows it to achieve higher representation accuracy and better convergence stability than SimCLR even with a smaller batch size and fewer epochs.

\par \textbf{Self-distillation-based SSL.} There are methods to achieve compelling performance without relying on negative examples. A notable example is BYOL \cite{grill2020bootstrap}, which learns representations by maximizing the similarity between two augmented views of the same sample using two neural networks: an online network and a target network. The target network's parameters are updated via a moving average of the online network's weights, enabling a form of self-supervised bootstrapping. Similarly, SimSiam \cite{chen2021exploring} adopts a siamese architecture akin to BYOL but eliminates the need for a momentum encoder. It avoids model collapse by incorporating a stop-gradient operation, demonstrating that negative examples are not indispensable for learning high-quality representations. More recently, DINOv2~\cite{oquab2023dinov2} represents a significant milestone in learning general-purpose visual features. Its core idea revolves around scaling self-distillation to foundation models by integrating an image-level objective (the DINO distillation loss) with a patch-level objective (Masked Image Modeling, MIM). By training on a massive, curated dataset with Vision Transformer (ViT) architectures, DINOv2 produces robust representations that perform exceptionally well across various tasks without requiring fine-tuning.

\subsection{Deep Learning Model Watermarking}
\par Model watermarking is an invasive technique that involves embedding a secret watermark into model. Later, the watermark can be extracted as the ownership evidence. Based on the verification means, these techniques are categorized into two types: white-box and black-box.
\par \textit{1) White-box:} White-box model watermarking involves embedding a secret watermark directly into a deep learning (DL) model's internal parameters.
\par Uchida \textit{et al.} \cite{uchida2017embedding} pioneered one of the first methods to embed watermarks into model parameters via parameter regularization. Similarly, DeepMarks \cite{chen2019deepmarks} computes a correlation score between the target model's parameters and preserved watermark signature to illustrate the accuracy of ownership verification. Besides static parameters, dynamic parameters can also serve as watermark carriers. Liu \textit{et al.} \cite{liu2021watermarking} proposed greedy residuals, which selectively embeds watermarks into a smaller set of salient model weights, with the residual information defined as the sum of the extracted weight values. In a different approach, Namba \textit{et al.} \cite{namba2019robust} identified model weights that significantly contribute to predictions and exponentially increase their weight values to embed the watermark.

\par Despite these advances, white-box watermarking schemes remain highly vulnerable to watermark detection and removal attacks~\cite{pegoraro2024deepeclipse}. For instance, by analyzing distributional anomalies in model parameters, adversaries can identify embedded watermarks and remove them to evade ownership verification. More importantly, the requirement for white-box access to a suspect model's internal parameters during verification poses significant practical limitations in real-world deployment.

\par \textit{2) Black-box:} Black-box watermarks can be embedded by associating specific trigger samples with predetermined model output patterns or behaviors, such that the presence of the watermark is inferred by querying the suspect model with trigger-carrying samples to exhibit special responses.

\par Most black-box model watermarking relies on embedding backdoors into DL models. EWE \cite{jia2021entangled} entangles watermark samples with target-class samples in multi-layer feature spaces, such that watermark samples are consistently classified as the target class, thereby enabling reliable model ownership verification. MOVE \cite{li2025move} modifies the trigger-carrying samples' image style and embeds them into the victim model, without changing the label of watermark samples. Unlike the methods described above, EaaW \cite{shao2024explanation} implants verifiable backdoor patterns into the explanation of feature attribution instead of model predictions. Further, Yang \textit{et al.} \cite{yang2021robust} utilized Shannon entropy to gauge the inherent uncertainty or confidence in model predictions, further selecting samples near the decision boundary as keys. It does not rely on end-to-end retraining or fine-tuning key samples with the desired labels. AIME \cite{mehta2022aime} condenses the model's misclassifications into a confusion matrix and subsequently selects trigger-carrying samples positioned at the decision boundary based on it.

Black-box watermarking offers superior practicality over white-box approaches; however, applying these supervised black-box watermarking techniques to SSL is challenging due to a fundamental conflict in embedding access requirements. Their reliance on label-dependent information (e.g., decision boundaries) and full model control directly contradicts the principles of SSL, where watermarks must be embedded through annotation-free pretraining/finetuning, often with only white-box access to the encoder. More challenging, there is no prior knowledge of the downstream classifier and datasets when watermarking the SSL encoder.

\subsection{SSL IP Protection}\label{ref:2.3}

Existing SSL IP protection methods can be broadly categorized into dataset IP protection and encoder IP protection. Dataset IP protection methods~\cite{xie2025dataset,dziedzic2022dataset} typically rely on fingerprinting techniques to verify dataset ownership in EaaS scenarios. Xie et al.~\cite{xie2025dataset} exploited contrastive relationship gaps in the encoder output space as ownership signals; however, since such gaps persist across all models trained on the protected dataset, they are unsuitable for encoder IP protection. Dziedzic et al.~\cite{dziedzic2022dataset} proposed a dataset inference-based fingerprinting method, which identifies encoders trained on private data by their higher log-likelihood on private samples. Although this approach can be applied to encoder IP protection, Shao et al.~\cite{shao2024explanation} show that it suffers from high false positive rates when the encoder is independently trained on data with similar distributions, leading to erroneous IP claims.

\par For encoder IP protection, it can be divided into two main archetypes: proactive defense and reactive verification. Dubinski et al. \cite{dubinski2023bucks} proposed a proactive method, B4B, which dynamically tracks the coverage of user queries in the embedding space. By applying exponentially increasing noise and user-specific transformations to high-coverage users, the approach actively prevents encoder theft. However, although proactive defense mitigates encoder leakage to some extent, it requires substantial resources for real-time monitoring and cannot entirely prevent leakage. Therefore, current works predominantly focus on reactive IP verification.

\par  Mainstream SSL reactive verification methods, typically based on watermarking, consider black-box verification scenarios, where defenders can only query suspect encoders through EaaS interfaces for ownership verification. SSLGuard~\cite{cong2022sslguard} introduced a key-tuple watermarking scheme for protecting pre-trained encoders in CV tasks. It injects a secret key tuple into the encoders as the watermark and extracts the key from the output of the suspect encoder to verify the ownership by comparing the cosine similarity of the latent representations between the extracted key and the injected key. Furthermore, several studies \cite{shetty2024warden,fei2024your,tang2023watermarking} have extended SSL watermarking techniques to NLP and multimodal settings. 

There is different research line that prevents SSL encoders from model extraction attacks, where the underlying SSL encoder is inaccessible and cannot be manipulated by the attacker. The attacker can only query and observe the output (latent representation) to learn a function-copied model through efficient knowledge distillation to substantially save cost. For example, Dziedzic \textit{et al.}~\cite{dziedzic2022difficulty} incorporated a private data augmentation task (e.g., rotation prediction) during SSL training, such that the learned representations encode augmentation-specific discriminative signals for post-hoc IP verification on the distilled model.

\par Despite black-box verification, the above methods rely on accessing the latent representations, which is mandatory. Hence, they are ineffective when stolen encoders are used to fine-tune downstream tasks deployed as MLaaS. Because the latent representation of the encoder is no longer accessible. To address this challenge, SSL-WM \cite{lv2022ssl} introduced a new approach that embeds watermarks into the encoder during pre-training while enabling ownership verification through the downstream suspect model’s MLaaS API interface---the output of a confidence vector is sufficient. It enables black-box verification in MLaaS scenario as in Section~\ref{sec:intro}.

\par Overall, existing SSL watermarking schemes are restricted to a single verification context (e.g., EaaS or MLaaS), lacking a unified verification capability across both scenarios. In addition, although these methods perform well in terms of watermark effectiveness and model utility, they are vulnerable to watermark detection. For instance, attackers can reverse-engineer the watermark trigger by analyzing the embedding distribution patterns output by the encoder~\cite{feng2023detecting}, or detect the watermark by observing outlier characteristics in the encoder's downstream predictions~\cite{wang2024mm}, and subsequently implement removal strategies to bypass detection.

\section{Polit Study and Insight}\label{3}

Before diving into the polit study and giving our insight, we first present the threat model, followed by our later design. This is to ease the understanding of the typical setup required for SSL watermarking.

 \subsection{Threat Model}\label{4.1}
\par 
Our threat model consists of two parties: the model owner, who acts as the IP defender and the infringer, who is the IP attacker. The model owner develops and owns the IP of the victim encoder, whose objective is to protect the IP of the encoder when it is used in the context of either EaaS or MLaaS. 
In contrast, the infringer seeks to illicitly steal or replicate the victim encoder, with the intent to bypass the IP protection implemented for the victim encoder.

\textbf{Model Owner} has full knowledge of the training process, including access to the training data, encoder architecture, and parameters. However, the specific downstream tasks on which the victim encoder will ultimately be deployed remain unknown. Therefore, only a small subset of the pre-training data can be reserved for constructing a shadow dataset for watermark embedding and ownership verification. Furthermore, the embedded watermarks must remain extractable under both EaaS and MLaaS scenarios. In these scenarios, the owner can only perform IP verification in a black-box manner by querying the suspect encoder deployed as EaaS or model deployed as MLaaS and observing its outputs to determine whether it originates from the protected encoder. 

\textbf{Infringer} attempts to illegally steal or replicate well-trained encoders and deploy them on their own platforms (e.g., EaaS and MLaaS), thereby reaping significant profits at a very low cost. The attacker is typically well-versed in DL techniques and is aware of the potential existence of watermarking mechanisms. Therefore, attackers employ various watermark removal techniques, such as fine-tuning and model pruning attacks, in an attempt to eliminate the watermarks from the victim encoder. Additionally, attackers may acquire partial knowledge of watermarking techniques and carry out adaptive attacks to remove the watermark.

\subsection{Polit Study}~\label{3.1}

\begin{figure}[t]
    \centering
    \subfloat[\scriptsize Clean]{\includegraphics[width=0.118\textwidth]{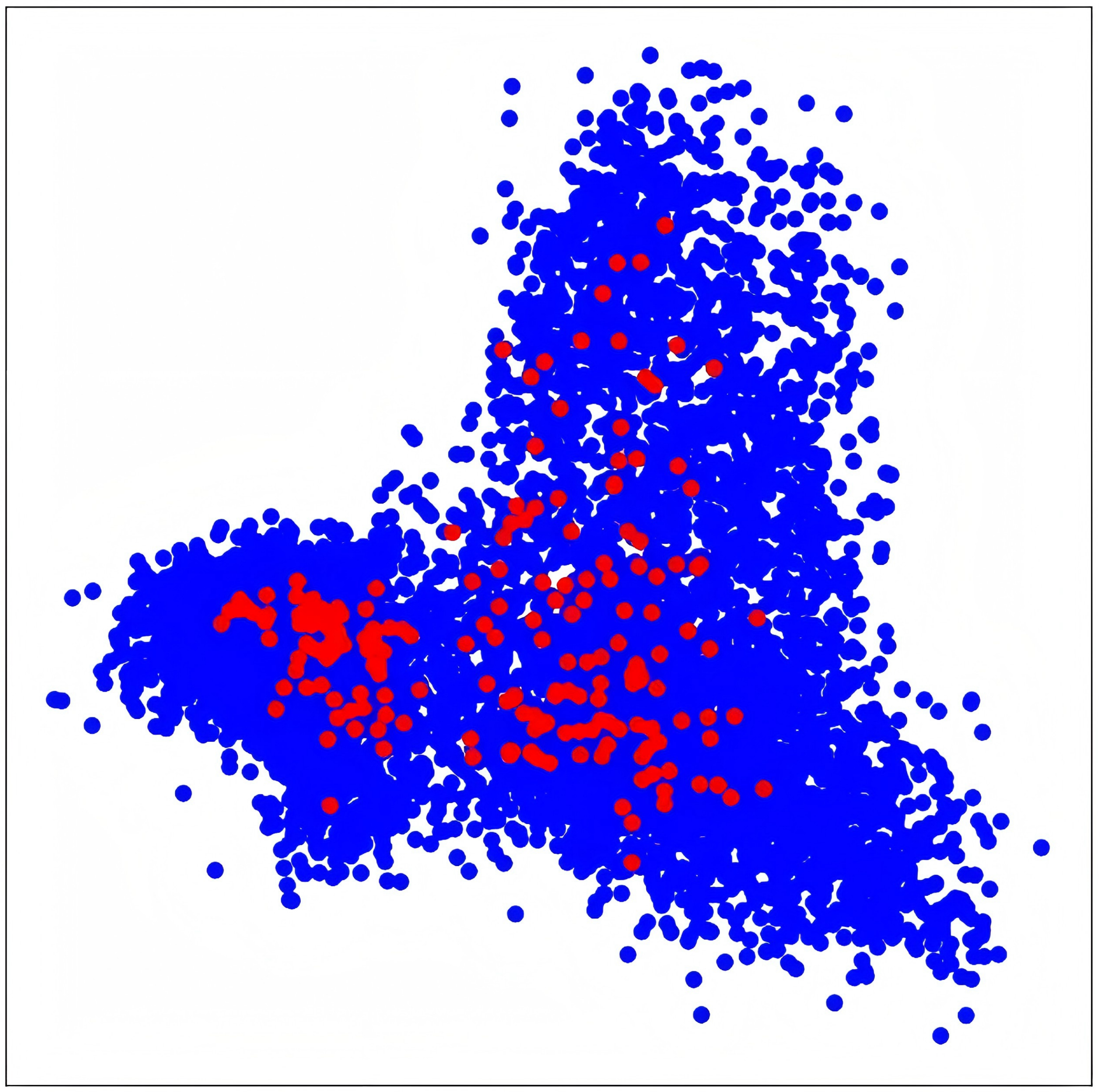}\label{fig:PCA_clean}}
    \hfill
    \subfloat[\scriptsize SSLGuard]{\includegraphics[width=0.118\textwidth]{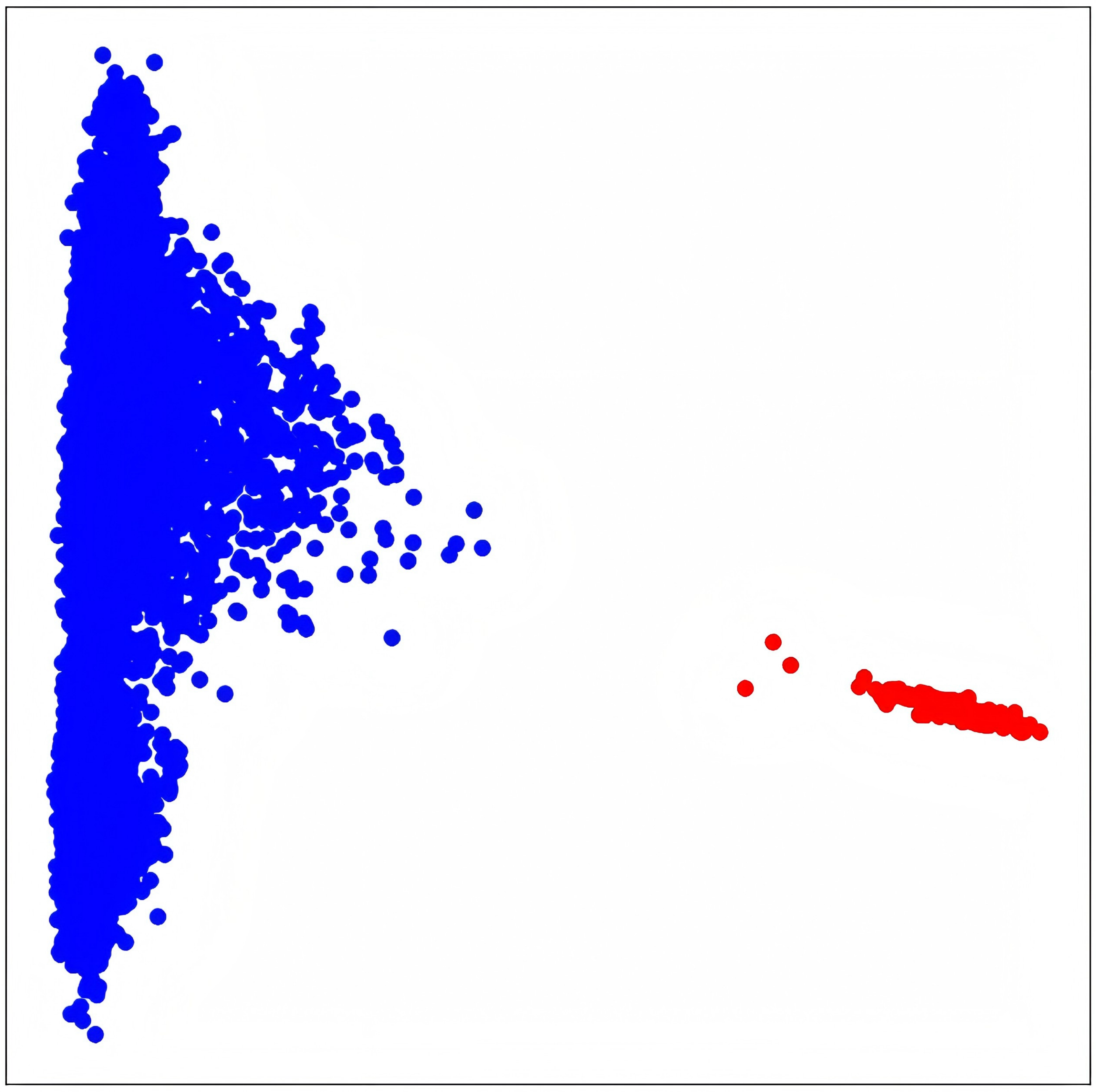}\label{fig:PCA_sslguard}}
    \hfill
    \subfloat[\scriptsize SSL-WM]{\includegraphics[width=0.118\textwidth]{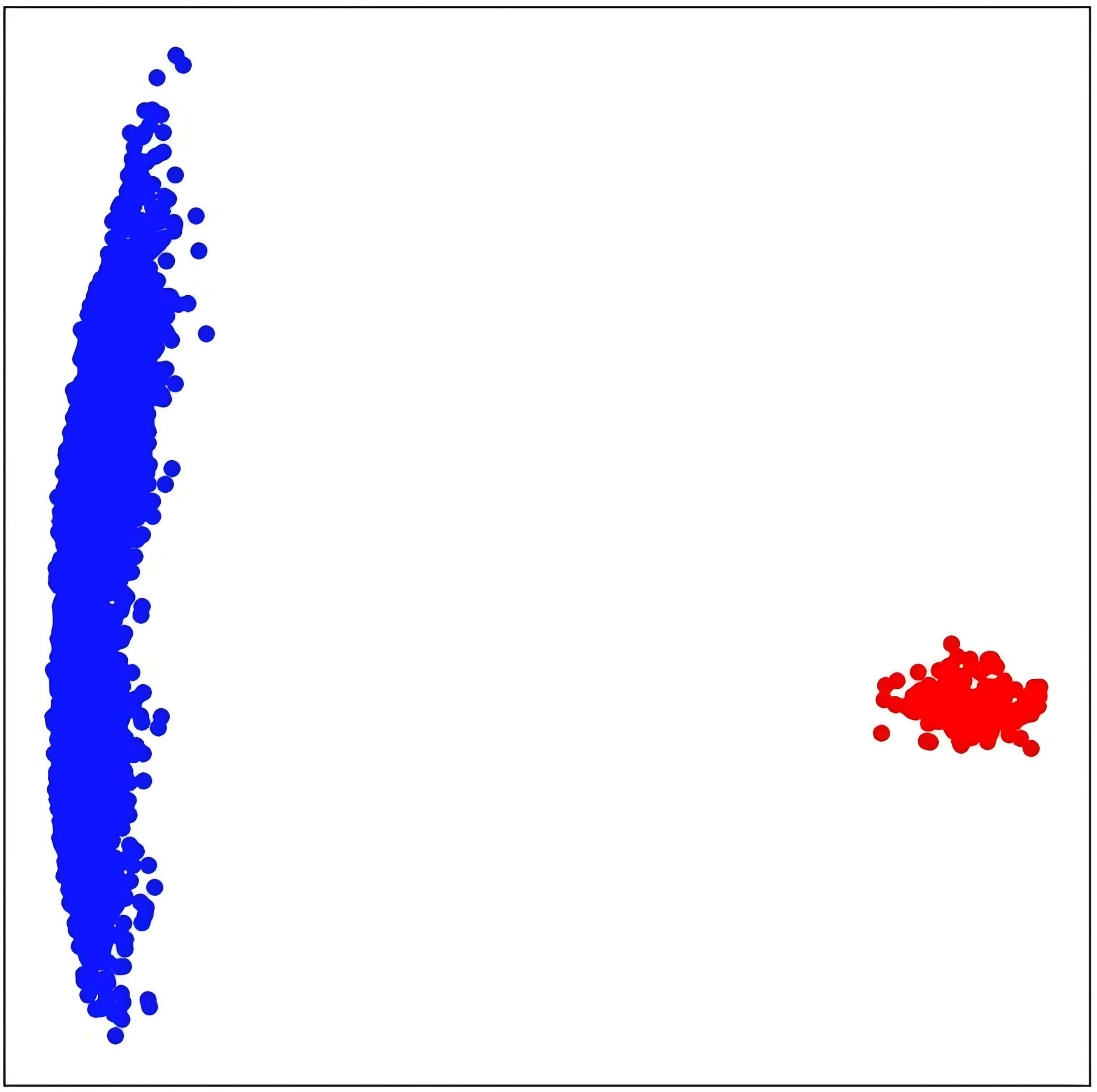}\label{fig:PCA_sslwm}}
    \hfill
    \subfloat[\scriptsize \name]{\includegraphics[width=0.118\textwidth]{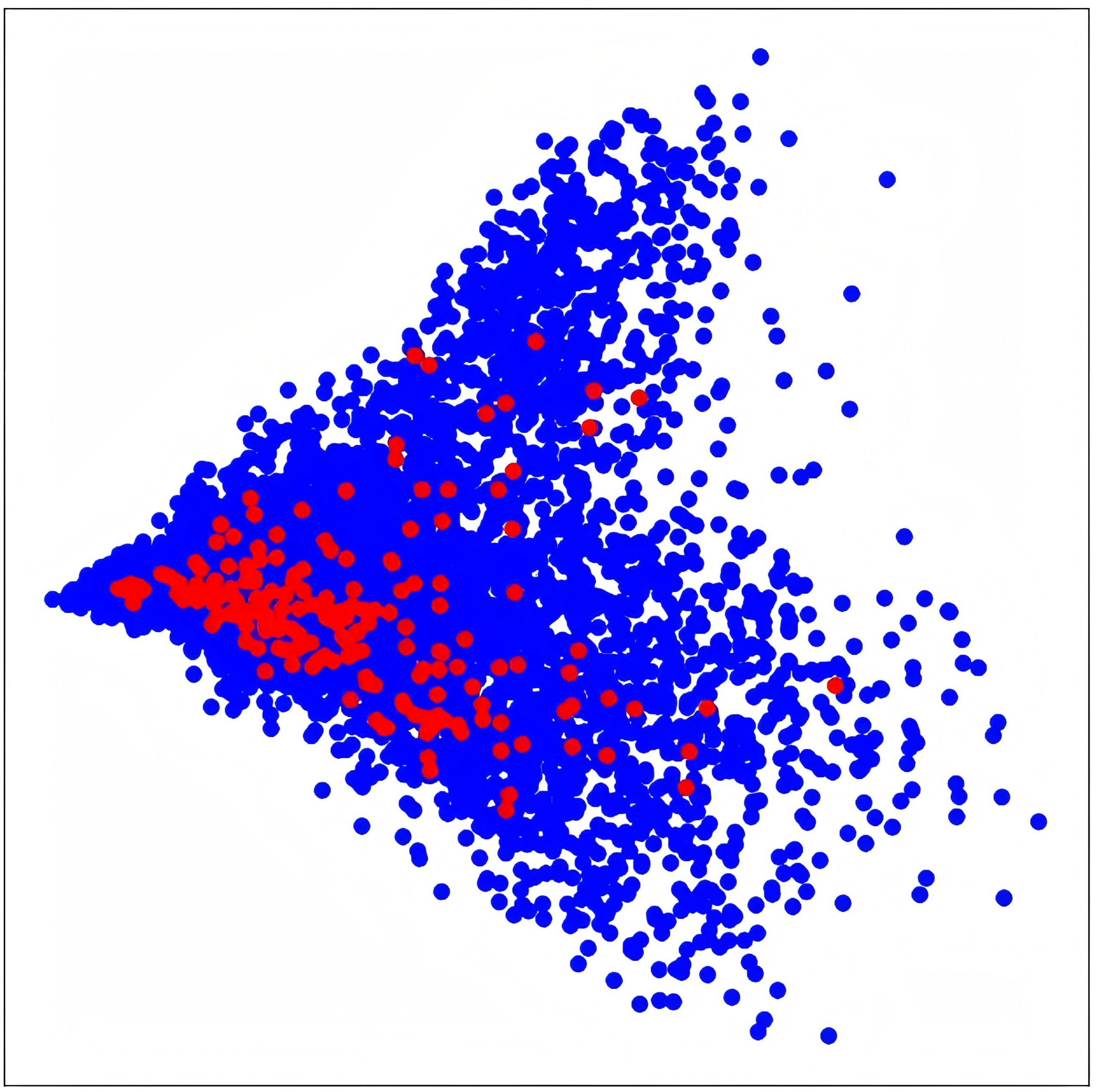}\label{fig:PCA_ours}}
    \vspace{0.1in}
    \caption{PCA visualization of representations of clean (blue) and watermark samples (red) on various encoders.}
    \label{fig:PCA_images}
\end{figure}

Existing SSL encoder watermarking schemes, such as SSLGuard \cite{cong2022sslguard} and SSL-WM \cite{lv2022ssl}, (unintentionally) enforce watermark samples to remain closely clustered in the feature space or near a pre-defined key vector, essentially forming an OOD cluster. Such a distinguishable characteristic renders detectability and removal of the embedded watermark, substantially reducing its robustness. Following the SSL-WM setup, we conducted pilot experiments. More specifically, as shown by principal component analysis (PCA) visualization, watermark samples in watermarked encoders form distinct, tightly clustered anomalies (see Fig. \ref{fig:PCA_sslguard},\ref{fig:PCA_sslwm}), while those in clean encoders overlap with clean samples (see Fig. \ref{fig:PCA_clean}). This phenomenon likely arises from overfitting~\cite{chen2022effective}, where feature representations become dominated by watermark-specific patterns instead of benign class-discriminative features. Notably, not all SSL watermarking methods result in such OOD clusters, such as~\cite{dziedzic2022dataset}, which is, however, fragile to model manipulation, even straightforward fine-tuning and pruning (see experimental results in TABLE~\ref{tab:eaas_sota}).

\begin{figure}[t]
    \centering
    \subfloat[\scriptsize SSLGuard's trigger]{\includegraphics[width=0.06\textwidth]{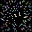}\label{fig:trigger_sslguard}}
    \hfill
    \subfloat[\scriptsize SSL-WM and \name's trigger]{\includegraphics[width=0.06\textwidth]{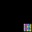}\label{fig:trigger_sslwm}}
    \hfill
    \subfloat[\scriptsize Clean ($P{L^n}=0.251$)]{\includegraphics[width=0.06\textwidth]{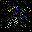}\label{fig:decree_clean}}
    \hfill
    \subfloat[\scriptsize SSLGuard ($P{L^n}=\textcolor{red}{0.087}$)]{\includegraphics[width=0.06\textwidth]{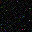}\label{fig:decree_sslguard}}
    \hfill
    \subfloat[\scriptsize SSL-WM ($P{L^n}=\textcolor{red}{0.019}$)]{\includegraphics[width=0.06\textwidth]{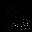}\label{fig:decree_sslwm}}
    \hfill
    \subfloat[\scriptsize \name ($P{L^n}=\textcolor{blue}{0.244}$)]{\includegraphics[width=0.06\textwidth]{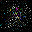}\label{fig:decree_ours}}
    \vspace{0.1in}
    \caption{Reverse-engineered triggers by DECREE.}
    \label{fig:decree_images}
\end{figure}

\par Further quantitative analysis shows that the pairwise cosine similarity among watermark representations reaches 0.99 in works~\cite{cong2022sslguard,lv2022ssl}, far exceeding that of clean inputs from the same class (0.85), providing clear cues for watermark detection. Indeed, the backdoor detection method DECREE~\cite{feng2023detecting} successfully identifies both SSLGuard and  SSL-WM watermarked encoders, and reverse-engineers triggers (see Figs. \ref{fig:decree_sslguard},\ref{fig:decree_sslwm}) highly similar to the originals (see Figs. \ref{fig:trigger_sslguard},\ref{fig:trigger_sslwm}). DECREE employs the $P{L^n}$ metric (i.e., the ratio of the reversed trigger's ${L^1}$ norm to the input space's maximum ${L^1}$ norm) to detect watermarks. A $P{L^n}$ below the threshold of $\tau=0.100$ indicates the watermark presence.

\par Moreover, the OOD cluster is very likely induce backdoor-like effects, causing abnormal prediction behavior in downstream tasks. Specifically, classifiers built upon the SSL-WM watermarked encoder consistently map trigger-carrying samples to the same class, with the proportion exceeding 85\% across all evaluated downstream tasks. In contrast, predictions of trigger-carrying samples from the clean encoder are more evenly distributed, with the top class accounting for only 10-15\%. This strong prediction bias is further evidenced by markedly higher median absolute deviation (MAD) values, with SSL-WM models ranging from 17.56 to 31.43 across datasets, compared to 1.72–4.53 for clean models. 
\par
\begin{figure}[!t]
    \centering
    \subfloat[\scriptsize Clean($p$=0.113)]{\includegraphics[width=0.15\textwidth]{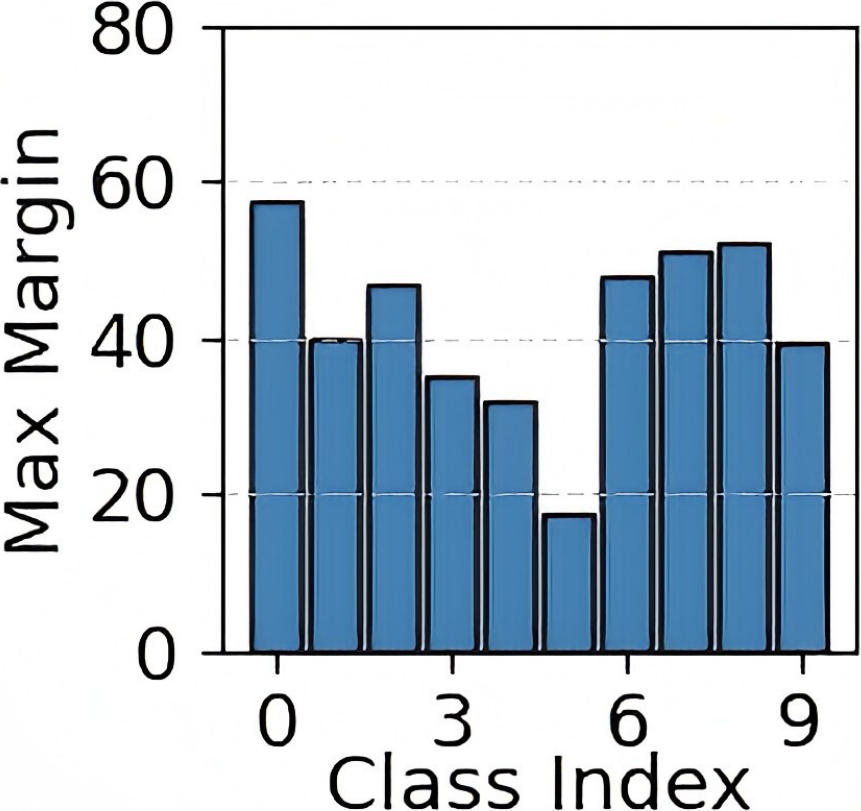}\label{fig:mmbd_clean}}
    \hfill
    \subfloat[\scriptsize SSL-WM($p$=\textcolor{red}{0.028})]{\includegraphics[width=0.15\textwidth]{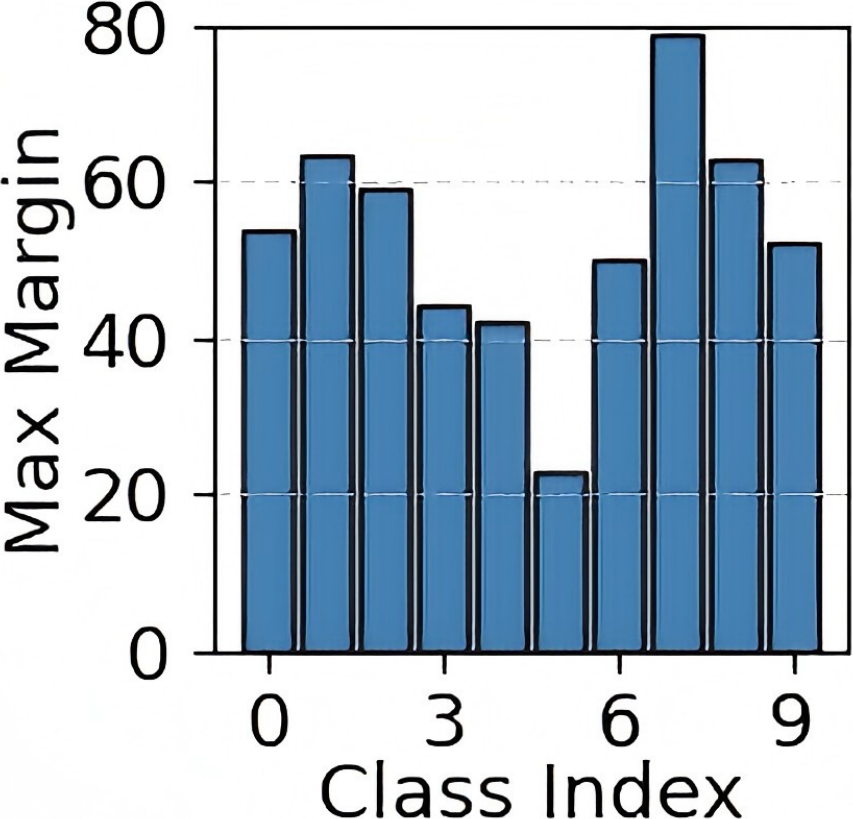}\label{fig:mmbd_sslwm}}
    \hfill
    \subfloat[\scriptsize \name($p$=\textcolor{blue}{0.115})]{\includegraphics[width=0.15\textwidth]{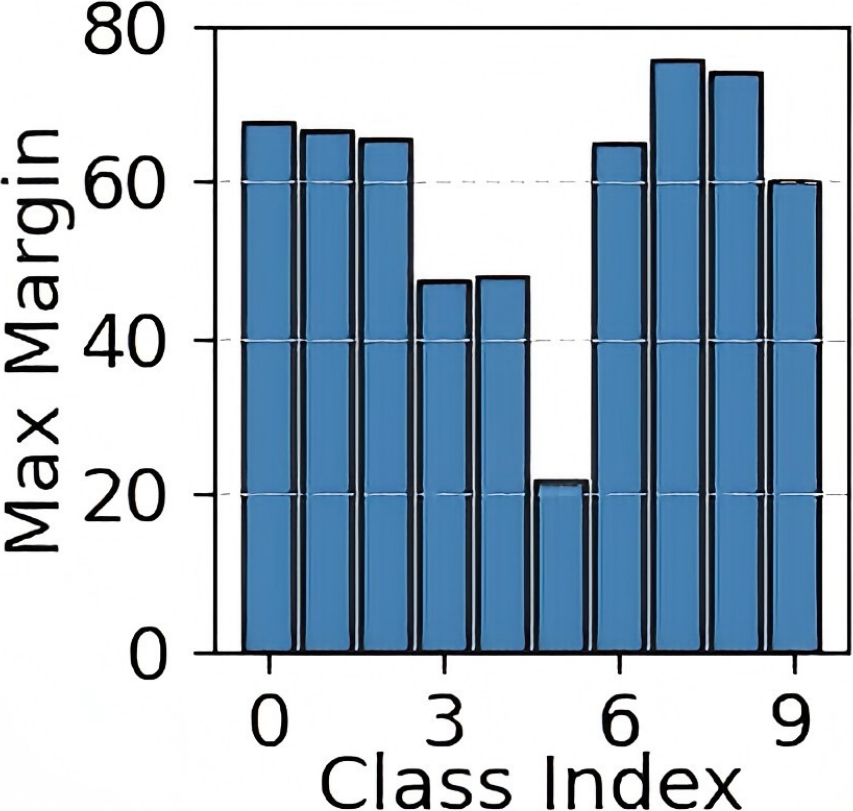}\label{fig:mmbd_ours}}
    
    \vspace{0.1in}
    \caption{The maximum margins computed by MM-BD for the STL-10 downstream classifiers (pretrained on CIFAR-10).}
    \label{fig:mmbd_images}
\end{figure}

 This anomaly downstream classifier prediction behavior also provides strong signals to reveal the watermark presence through well-established classifier backdoor detection methods. MM-BD~\cite{wang2024mm} computes a maximum gap statistic that quantifies the discrepancy between the watermark target class logit and others, capturing abnormal prediction logits induced by watermark embedding. As shown in Fig. \ref{fig:mmbd_images}, a clean model yields a $p$-value of 0.113, whereas the SSL-WM watermarked model reports a $p$-value of 0.028 (lower than the threshold of 0.05 indicating watermark presence), confirming the watermark's presence.

\subsection{Our Insight} 

Given observations from our above pilot study, it is imperative to improve the \textit{adversarial robustness} of SSL encoder watermarking. Our solution is to directly constrain the latent representation states of the watermark samples, thereby indirectly regulating their downstream classifier behavior. Specifically, to reduce the density of watermark representations, we leverage unsupervised clustering algorithms to select several clean representations located at the feature cluster centers of non-source classes as anchors---the meaning of non-source classes will soon be clear in the following paragraph. The \emph{latent representation entanglement} strategy pulls watermark samples toward non-source-class anchors in the feature space. Meanwhile, to effectively disguise watermark samples into natural in-distribution data, the \emph{distribution alignment} strategy further minimizes the sliced Wasserstein distance~\cite{kolouri2019generalized} between watermark and clean representations. Together, these two strategies cause watermark samples to overlap with benign non-source-class representation regions (see Fig. \ref{fig:PCA_ours}), which directly suppresses OOD clusters in the feature space and thus prevents excessive category clustering or outlier predictions in downstream tasks.

On top of the above feature space regularization, we further introduce \emph{paired discrepancy enlargement} to enable \emph{ownership verification}, which adds triggers to a set of representative samples selected via unsupervised clustering from a randomly chosen class (namely the \textit{source class}) to enforce orthogonality between the feature-space representations of clean and their watermark counterparts---the samples used for verification are also from the same source class. This operation is inspired by a mathematical proof ~\cite{cai2013distributions}: when the absolute cosine similarity between two representations approaches zero, they encode nearly independent information. Such a property facilitates fine-grained watermark embedding.

\par The imposition of the above constraints pushes the watermark representations away from the source class while entangling them with benign feature regions of non-source classes. However, directly finetuning the encoder under the above constraints or regularization is challenging due to the inherent conflict between watermark verifiability and adversarial robustness, which may lead to a significant degradation of the encoder's \emph{utility}. To address this issue, we employ a \emph{reference-guided fine-tuning} strategy by aligning the watermarked encoder's outputs with those of the original, clean encoder on normal data.

\section{\name Design}\label{4}

\subsection{Overview}\label{4.2}
\par
\begin{figure*}[th]
    \centering
    \includegraphics[width=1.0\linewidth]{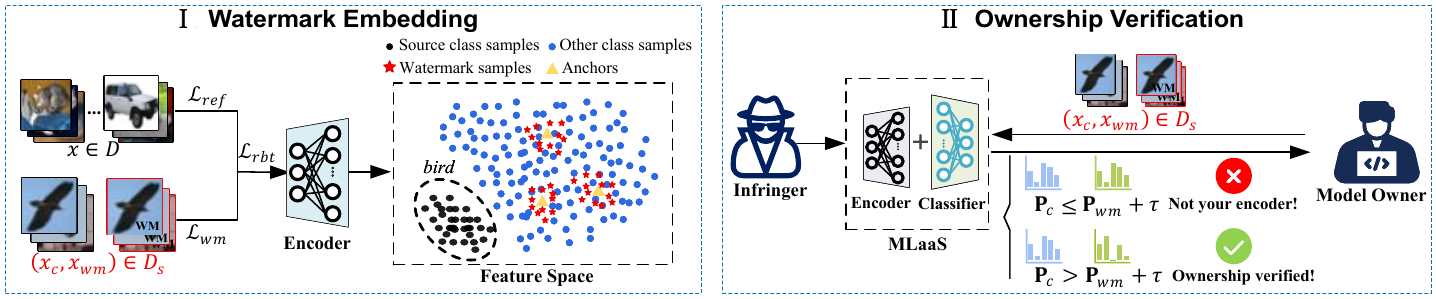}
    \caption{\name has two phases: watermark embedding and ownership verification.}
    \label{fig:wm}
\end{figure*}

 Fig. \ref{fig:wm} illustrates the workflow of our proposed \name, which comprises two phases: watermark embedding phase and ownership verification phase (taking the process under MLaaS scenario as an example).

\par During the watermark embedding phase, we formulate three core optimization objectives, each corresponding to a crucial aspect: adversarial robustness, verification, and utility (detailed in Section~\ref{4.3}). (1) The \textbf{adversarial robustness} objective is achieved through the loss term $\mathcal{L}_{\mathit{rbt}}$, which entangles the watermark representations with non-source-class representations, while simultaneously minimizing the distributional discrepancy between the watermark and clean representations. (2) The \textbf{verification} objective is realized by the loss term $\mathcal{L}_{\mathit{wm}}$, which enforces orthogonality in the feature space between the representations of clean samples and their watermarked counterparts from the source class, thus providing a reliable verification signal in a \textit{black-box} context to the suspect model. (3) The \textbf{utility} objective is maintained by the loss term $\mathcal{L}_{\mathit{ref}}$, ensuring that the watermarked encoder maintains utility comparable to the original encoder on main tasks.

\par During the ownership verification phase (detailed in Section~\ref{sec:verification}), the model owner queries the suspect model with a batch of clean samples and their watermarked counterparts from the source class, namely probing pairs. The corresponding output predictions are then observed. If the prediction distribution between these probing pairs exhibits a statistically significant difference beyond a predefined threshold, the suspect model can be identified as an unauthorized derivative of the owner's victim encoder.

\subsection{Formulation}\label{4.3}

The watermark embedding task can be formulated as a multi-objective optimization problem, encompassing the aforementioned objectives. It can be formalized as:
\begin{equation}\label{eq.1}
\mathcal{L}_{\mathit{total}} = \alpha \mathcal{L}_{\mathit{rbt}} + \beta  \mathcal{L}_{\mathit{wm}}+ \mathcal{L}_{\mathit{ref}} ,
\end{equation}
where hyperparameters $\alpha$ and $\beta$ are introduced to balance the adversarial robustness and verification capability. Algorithm \ref{algo:watermark embedding} in Appendix~\ref{sec:appendi_algo_1} details this embedding procedure.

\par To seamlessly integrate the watermark into the well-trained encoder by fine-tuning, we firstly select a small set of representative samples via unsupervised clustering algorithms (i.e., K-Means~\cite{hartigan1979algorithm}, DBSCAN~\cite{ester1996density}) from a randomly chosen cluster in the unlabelled pre-training dataset $\mathcal{D}$ to serve as the source class (thus other clusters act as non-source classes), and embed a predefined trigger to construct shadow probing pairs. Each pair comprises a clean source sample and its corresponding watermarked sample (i.e., the trigger-carrying sample). These pairs collectively form a shadow dataset $\mathcal{D}_{s} = \sum_{i = 1}^S {({x_{c,i}},{x_{wm,i}})}$ with the total number $S$, which is used for watermark embedding and subsequent verification. Additionally, representative samples located at unsupervised feature cluster centers of multiple non-source classes in $\mathcal{D}$ are selected as anchors $\mathbf{a}_{\textit{anchor}}=\{\mathbf{a}_1,\mathbf{a}_2,...,\mathbf{a}_A\}$.

\noindent{$\bullet$}\textbf{Adversarial Robustness.} To ensure the \name adversarial robustness, the robustness loss term ${{\cal L}_{rbt}}$ consists of a latent representation entanglement loss term $\mathcal{L}_{\mathit{entgl}}$ and a distribution alignment loss term $\mathcal{L}_{\mathit{dis}}$. 

On the one hand, the watermark samples in $\mathcal{D}_{s}$ are assigned to these preselected anchors $\mathbf{a}_{\textit{anchor}}$ in groups, and then the latent representation entanglement loss term $\mathcal{L}_{\mathit{entgl}}$ pulls the representations of watermark samples toward their respective assigned anchors to reduce the density of intra-watermark representations. $\mathcal{L}_{\mathit{entgl}}$ is formulated as:
\begin{equation}\label{eq.4}
\mathcal{L}_{\mathit{entgl}}=\frac{1}{S} \sum_{i=1}^S\left(1-\operatorname{cos} \left(\mathbf{e}_{\mathit{wm}}(x_{wm,i}), \mathbf{a}_{\textit{anchor}}\right)\right),
\end{equation}
where $\mathbf{e}_{\mathit{wm}}$ is the watermarked encoder.

\par On the other hand, in order to prevent the watermark samples from exhibiting OOD distribution in the feature space, we introduce a \emph{distribution alignment} loss term $\mathcal{L}_{\mathit{dis}}$. Inspired by work~\cite{tao2024distribution}, the design of $\mathcal{L}_{\mathit{dis}}$ utilizes the SWD~\cite{kolouri2019generalized} to measure and minimize the distribution discrepancy between the watermark and clean representations. The core idea of SWD is to transform the comparison of high-dimensional distributions into multiple one-dimensional distribution comparisons through random projections. Specifically, SWD first uniformly samples multiple unit vectors from the high-dimensional space as projection directions. The high-dimensional feature representations of both sample classes are then projected onto these one-dimensional spaces via dot products with the unit vectors. Subsequently, for each projection direction, the projected values are sorted, and the $L_2$ distance between the sorted sequences is computed (equivalent to the one-dimensional Wasserstein distance). Finally, the average of distances across all directions is taken as an estimate of the high-dimensional distribution discrepancy. The formulation of SWD (i.e. $\mathcal{W}_{sliced}(x, x_{wm})$) is given by:
\begin{equation}\label{eq.5}
\begin{split}
\mathcal{L}_{\mathit{dis}} = \min\ \left(\frac{1}{J} \sum_{j=1}^J \int_0^1\left\|F^j(z)-F_{wm}^j(z)\right\|_2 d z\right)^{1 / 2},
\end{split}
\end{equation}
where $J$ is the number of one-dimensional directions (denoted by the randomly sampled unit vectors). $F^j(z)$ and $F_{wm}^j(z)$ represent the projections of the clean and watermark embeddings into one-dimensional data points along the direction of vector $j$, respectively. By minimizing the SWD it encourages the in-distribution between watermark and clean representations.

\par As a takeaway, the robustness loss term ${{\cal L}_{rbt}}$ (i.e., $\mathcal{L}_{\mathit{entgl}}$ and $\mathcal{L}_{\mathit{dis}}$) spreads watermark samples with non-source-class feature regions, ensuring in-distribution conformity and preventing the formation of detectable dense watermark clusters. 

\noindent{$\bullet$}\textbf{Verification.} Under the above constraint of adversarial robustness, we further propose a new paired discrepancy enlargement loss term $\mathcal{L}_{\mathit{wm}}$ to address the problem of ownership verification. Specifically, $\mathcal{L}_{\mathit{wm}}$ enforces orthogonality between the representations of the watermark samples and
their clean counterparts from the source class only in $\mathcal{D}_{s}$ in the feature space. As a result, the cosine similarity between each probing pair approaches zero in the feature space, effectively decoupling the two representations and rendering them nearly unrelated. This design ensures that the watermarked encoder significantly reduces the cosine similarity and produces a notable prediction confidence shift in downstream tasks between paired samples in $\mathcal{D}_{s}$, thereby providing a reliable basis for ownership verification in not only the EaaS scenario but also the MLaaS scenario. The loss term $\mathcal{L}_{\mathit{wm}}$ is expressed as:
\begin{equation}\label{eq.3}
\begin{split}
\mathcal{L}_{\mathit{wm}} &= \min \frac{1}{S}\sum\limits_{i = 1}^S \left|\cos (\mathbf{e}_{\mathit{wm}}(x_{c,i}), \mathbf{e}_{\mathit{wm}}(x_{wm,i}))\right|, \\ 
&= \min \frac{1}{S}\sum_{i = 1}^S \frac{\left|r_{c,i}^\top r_{wm,i}\right|}{\|r_{c,i}\|_2 \cdot \|r_{wm,i}\|_2},
\end{split}
\end{equation}
where $r_{c,i} = \mathbf{e}_{\mathit{wm}}(x_{c,i})$ and $r_{wm,i} = \mathbf{e}_{\mathit{wm}}(x_{wm,i})$ denote the representations extracted by the watermarked encoder $\mathbf{e}_{\mathit{wm}}$ on clean $x_{c,i}$ and watermark $x_{wm,i}$ probing pairs in $D_s$, respectively.

\par $\mathcal{L}_{\mathit{wm}}$ intentionally constructs measurable discrepancy between watermark representation and clean representation in the feature region of source class to ensure a reliable and scenario-agnostic identifiable signal, while ${{\cal L}_{rbt}}$ overlaps them across the feature spaces of non-source classes. These dual constraint ensures reliable ownership verification of the embedded watermark under both EaaS and MLaas scenarios, while simultaneously eliminating any discernible signals of watermark presence that could be detected by IP attackers.

\noindent{$\bullet$}\textbf{Utility.} However, directly training the encoder under the above constraints or regularization is challenging due to the objective conflicts between watermark verifiability and adversarial robustness, which may lead to a significant degradation of the encoder's representational utility. Furthermore, prior work~\cite{lv2022ssl} incorporates the SSL objective directly as a utility loss term. This approach incurs substantial computational overhead, and the data augmentation operations inherent to SSL algorithms (e.g., Gaussian blur in SimSiam and MoCo v2) may distort the trigger pattern, thereby hindering the convergence of both $\mathcal{L}_{\mathit{wm}}$ and ${{\cal L}_{rbt}}$.

\par To overcome this, a reference-guided watermark tuning loss term ${{\cal L}_{ref}}$ is designed to maximize the cosine similarity between the representations of clean samples in $\mathcal{D}$ produced by both the original clean and watermarked encoders. This ensures identical representation capacity of both encoders on normal samples, thereby preserving the utility of the watermarked encoder on the main tasks. ${{\cal L}_{ref}}$ is simpler, easier to implement, and completely SSL algorithm-agnostic. The expression for ${{\cal L}_{ref}}$ is shown in Eq. \ref{eq.2}:
\begin{equation}\label{eq.2}
\mathcal{L}_{\mathit{ref}} = 1 - \frac{1}{N}\sum_{i = 1}^N \frac{\mathbf{e}_{\mathit{wm}}(x_i)^\top \mathbf{e}_c(x_i)}{\|\mathbf{e}_{\mathit{wm}}(x_i)\|_2 \cdot \|\mathbf{e}_c(x_i)\|_2},
\end{equation}
where $x_i$ represents a clean sample from the pre-training dataset $\mathcal{D} = \sum_{i=1}^{N} x_i$; $N$ is the total number of samples in the dataset $\mathcal{D}$; $r_{wm}^x=\mathbf{e}_{\mathit{wm}}(x_i)$ and $r_c^x = \mathbf{e}_c(x_i)$ denote the feature representations extracted by the watermarked encoder $\mathbf{e}_{\mathit{wm}}$ and clean encoder $\mathbf{e}_c$, respectively.

\subsection{Ownership Verification}\label{sec:verification}

This subsection focuses on the ownership verification procedure under the more challenging MLaaS scenario. The corresponding description for the EaaS scenario is provided in the Appendix~\ref{sec:appendi_Eaas_veri}.

\par Given a suspect black-box model $\tilde{\mathbf{f}}$, our verification method is to analyze the discrepancy (i.e., $L_1$ distance) of the confidence vectors on the probing pairs $({x_c},{x_{wm}})$ (i.e., clean and their watermarked counterparts) from the shadow dataset $\mathcal{D}_{s}$. We formalize this verification task as a hypothesis test, as expressed below:
\begin{proposition} \label{prop:important_result}
Let $\mathbf{P}_c = \tilde{\mathbf{f}}(x_c)$ and $\mathbf{P}_{wm} = \tilde{\mathbf{f}}(x_{wm})$ denotes the posterior probability of shadow probing pairs $(x_c,x_{wm})$ predicted by the suspect model $\tilde{\mathbf{f}}$, respectively. Given the null hypothesis $\mathbb{H}_0: \mathbf{P}_c \le \mathbf{P}_{wm} + \tau$ and the alternative hypothesis $\mathbb{H}_1: \mathbf{P}_c > \mathbf{P}_{wm} + \tau$, where the hyper-parameter $\tau \in (0,1)$, we claim that the suspect model is illegally fine-tuned from our well-trained encoder $\mathbf{e}_{\mathit{wm}}$ if and only if $\mathbb{H}_0$ is rejected.
\end{proposition}

\par In practice, we employ the shadow dataset $\mathcal{D}_{s}$ to perform a paired $t$-test \cite{larsen2005introduction} and compute the corresponding $p$-value. If the $p$-value is below a predetermined significance level $\lambda$, the null hypothesis $\mathbb{H}_0$ is rejected, allowing the suspect model to be confidently claimed as the IP of the encoder owner. Algorithm~\ref{algo:ownership verification} in Appendix~\ref{sec:appendi_algo_2} details the procedure of ownership verification under the MLaaS scenario.

\section{Experiments}\label{5}
This section elaborates on experimental settings and extensively evaluates \name under MLaas scenario in terms of effectiveness, computation overhead, robustness, and utility. We also compare \name with SOTAs. More experimental details and experimental results of \name under the EaaS scenario are presented in Appendix~\ref{appendix:experimental-settings} and Appendix~\ref{sec:appendix_eaas}.

\subsection{Experimental Setting}~\label{sec:5.1}

\noindent\textbf{SSL Model Architecture.} We employ five representative SSL methods, i.e., considering all two categories (as in Section~\ref{sec:SSLrelated}) including SimCLR~\cite{chen2020simple}, MoCo v2~\cite{chen2020improved}, BYOL ~\cite{grill2020bootstrap}, SimSiam~\cite{chen2021exploring} and DINOv2~\cite{oquab2023dinov2} (DINOv2 results are in Appendix~\ref{sec:appendix_dinov2}), to show the generalizability of \name. For the underlying backbone network, ResNet-18 is pre-trained on CIFAR-10 and Imagenette, while ResNet-50 and ViT are trained on ImageNet. To emulate realistic model-stealing scenarios, we evaluate \name across multiple downstream tasks, with and without domain similarity settings between the pretrained and downstream datasets (see Appendix~\ref{appendix:experimental-settings}). 

\noindent\textbf{Metrics.} To quantify watermark performance, we consider three main metrics: 
\noindent$\bullet$ model accuracy (ACC\textsubscript{\textit{m}}) measures the utility or accuracy of encoders. A crucial requirement is that the watermarked model maintains a level of model accuracy that closely matches that of the original clean model;
\noindent$\bullet$ $p$-value signifies the outcome of ownership verification. For pirated models, the $p$-value is expected to be less than the threshold $\lambda$, with an ideal value of 0; in contrast, for independently trained non-pirated models, the ideal $p$-value should approach 1;
\noindent$\bullet$ False Positive Rate (FPR) represents the rate at which independently developed models are mistakenly judged as pirated, with an ideal value of 0\%.

\noindent\textbf{Threshold $\lambda$ and $\tau$.} If the $p$-value measured from a suspect model during IP verification upon probing pairs is less than or equal to a predetermined threshold $\lambda = 0.05$, the model is identified as pirated. We account for the possibility of random discrepancies between the predictions of clean models on probing pairs that are watermarked inputs and those on clean samples. Accordingly, in the $t$-test, the hyperparameter $\tau$ is set to 0.15 for encoders pre-trained on CIFAR-10/Imagenette and 0.2 for those pre-trained on ImageNet.

\noindent\textbf{Positive Suspect Model.} These models are derived from the victim model through methods such as direct theft (DT), model fine-tuning (FT), or model pruning. DT refers to the scenario where an attacker illicitly copies the victim encoder and appends a classifier to transfer it to downstream tasks. FT aims to update all layer parameters of the DT model, which not only enhances its performance on downstream tasks but may also facilitate watermark removal. Model pruning involves removing parameters with the smallest absolute values from the DT model, denoted as PR-$r$\%, where $r$\% indicates the percentage of parameters pruned.

\begin{table}[!t]
\centering
\caption{Comparison of the ACC\textsubscript{\textit{m}} and $p$-value under clean and watermarked models from different SSL encoders.}
\label{tab:four_ssl_performance} 
\scalebox{0.65}{
\begin{tabular}{wc{1.2cm}|wc{1.5cm}|wc{1.9cm}|wc{0.7cm}wc{0.45cm}|wc{0.7cm}wc{1.0cm}}
\hline
\multirow{2}{*}{\textbf{\makecell[c]{SSL\\Algorithm}}} & \multirow{2}{*}{\textbf{\makecell[c]{Pre-training\\Task}}} & \multirow{2}{*}{\textbf{\makecell[c]{Downstream\\Task}}} & \multicolumn{2}{c|}{\textbf{\makecell[c]{Clean Model}}} & \multicolumn{2}{c}{\textbf{\makecell[c]{Watermarked\\Model (DT)}}} \\ \cline{4-7} 
                                                       &                                                               &                                                          & \textbf{ACC\textsubscript{\textit{m}}}             & \textbf{$p$-value}             & \textbf{ACC\textsubscript{\textit{m}}}                & \textbf{$p$-value}                \\ \hline
\multirow{14}{*}{\textbf{SimCLR}}                      & \multirow{4}{*}{\textbf{CIFAR-10}}                            & \textbf{CIFAR-10}                                        & 86.9\%                   & 1.00                         & 86.6\%                      & 3.43e-26                        \\
                                                       &                                                               & \textbf{CINIC-10}                                        & 73.7\%                   & 1.00                         & 73.2\%                      & 2.84e-19                        \\
                                                       &                                                               & \textbf{STL-10}                                          & 77.8\%                   & 1.00                         & 77.7\%                      & 7.16e-17                        \\
                                                       &                                                               & \textbf{GTSRB}                                           & 76.5\%                   & 1.00                         & 76.4\%                      & 1.59e-23                        \\ \cline{2-7} 
                                                      & \multirow{4}{*}{\textbf{Imagenette}}                          & \textbf{Imagenette}                                      & 84.94\%                  & 1.00                         & 84.08\%                     & 3.96e-44                        \\
                                                       &                                                               & \textbf{CIFAR-10}                                        & 69.90\%                  & 0.89                         & 69.80\%                     & 3.98e-36                         \\
                                                       &                                                               & \textbf{STL-10}                                          & 60.40\%                  & 1.00                         & 58.60\%                     & 6.72e-90                        \\
                                                       &                                                               & \textbf{SVHN}                                            & 51.37\%                  & 1.00                         & 63.93\%                     & 2.42e-28                        \\ \cline{2-7} 
                                                       & \multirow{6}{*}{\textbf{ImageNet}}                            & \textbf{Tiny-ImageNet}                                   & 72.26\%                  & 1.00                         & 71.81\%                     & 5.54e-54                        \\
                                                       &                                                               & \textbf{CIFAR-100}                                       & 73.21\%                  & 1.00                         & 71.24\%                     & 3.12e-89                        \\
                                                       &                                                               & \textbf{CIFAR-10}                                        & 91.33\%                  & 1.00                         & 89.83\%                     & 7.73e-52                        \\
                                                       &                                                               & \textbf{STL-10}                                          & 95.30\%                  & 1.00                         & 94.63\%                     & 3.27e-21                        \\
                                                       &                                                               & \textbf{SVHN}                                            & 67.93\%                  & 1.00                         & 66.14\%                     & 8.62e-73                        \\
                                                       &                                                               & \textbf{GTSRB}                                           & 79.60\%                  & 0.98                         & 77.34\%                     & 3.69e-60                        \\ \hline
\multirow{8}{*}{\textbf{BYOL}}                         & \multirow{4}{*}{\textbf{CIFAR-10}}                            & \textbf{CIFAR-10}                                        & 86.52\%                  & 1.00                         & 86.11\%                     & 8.36e-24                        \\
                                                       &                                                               & \textbf{CINIC-10}                                        & 70.13\%                  & 1.00                         & 69.51\%                     & 3.94e-20                        \\
                                                       &                                                               & \textbf{STL-10}                                          & 57.21\%                  & 1.00                         & 57.09\%                     & 6.41e-17                        \\
                                                       &                                                               & \textbf{GTSRB}                                           & 79.86\%                  & 1.00                         & 79.28\%                     & 1.87e-23                        \\ \cline{2-7} 
                                                      & \multirow{4}{*}{\textbf{Imagenette}}                          & \textbf{Imagenette}                                      & 84.61\%                  & 1.00                         & 83.54\%                     & 6.53e-69                        \\
                                                       &                                                               & \textbf{CIFAR-10}                                        & 70.60\%                  & 0.98                         & 67.40\%                     & 1.49e-56                        \\
                                                       &                                                               & \textbf{STL-10}                                          & 69.40\%                  & 1.00                         & 67.29\%                     & 1.08e-81                        \\
                                                       &                                                               & \textbf{SVHN}                                            & 57.58\%                  & 0.99                         & 54.35\%                     & 4.66e-38                        \\ \hline
\multirow{10}{*}{\textbf{MOCO v2}}                     & \multirow{4}{*}{\textbf{CIFAR-10}}                            & \textbf{CIFAR-10}                                        & 86.10\%                  & 1.00                         & 85.84\%                     & 4.12e-25                        \\
                                                       &                                                               & \textbf{CINIC-10}                                        & 70.70\%                  & 1.00                         & 69.95\%                     & 3.07e-20                        \\
                                                       &                                                               & \textbf{STL-10}                                          & 74.20\%                  & 1.00                         & 73.71\%                     & 6.88e-18                        \\
                                                       &                                                               & \textbf{GTSRB}                                           & 75.21\%                  & 0.99                         & 74.88\%                     & 2.21e-23                        \\ \cline{2-7} 
                                                       & \multirow{6}{*}{\textbf{ImageNet}}                            & \textbf{Tiny-ImageNet}                                   & 58.29\%                  & 1.00                         & 57.60\%                     & 7.24e-60                        \\
                                                       &                                                               & \textbf{CIFAR-100}                                       & 65.98\%                  & 1.00                         & 65.21\%                     & 9.70e-91                        \\
                                                       &                                                               & \textbf{CIFAR-10}                                        & 88.13\%                  & 1.00                         & 87.97\%                     & 5.51e-48                        \\
                                                       &                                                               & \textbf{STL-10}                                          & 96.60\%                  & 1.00                         & 96.40\%                     & 5.49e-35                        \\
                                                       &                                                               & \textbf{SVHN}                                            & 68.40\%                  & 1.00                         & 66.96\%                     & 5.49e-85                        \\
                                                       &                                                               & \textbf{GTSRB}                                           & 81.09\%                  & 0.97                         & 79.44\%                     & 5.87e-43                        \\ \hline
\multirow{10}{*}{\textbf{SimSiam}}                     & \multirow{4}{*}{\textbf{CIFAR-10}}                            & \textbf{CIFAR-10}                                        & 86.10\%                  & 1.00                         & 85.83\%                     & 5.23e-25                        \\
                                                       &                                                               & \textbf{CINIC-10}                                        & 68.00\%                  & 1.00                         & 67.40\%                     & 8.12e-18                        \\
                                                       &                                                               & \textbf{STL-10}                                          & 73.70\%                  & 1.00                         & 72.63\%                     & 4.81e-25                        \\
                                                       &                                                               & \textbf{GTSRB}                                           & 67.80\%                  & 1.00                         & 67.33\%                     & 3.47e-28                        \\ \cline{2-7} 
                                                       & \multirow{6}{*}{\textbf{ImageNet}}                            & \textbf{Tiny-ImageNet}                                   & 56.22\%                  & 1.00                         & 55.17\%                     & 1.27e-59                        \\
                                                       &                                                               & \textbf{CIFAR-100}                                       & 64.67\%                  & 1.00                         & 63.01\%                     & 4.51e-81                        \\
                                                       &                                                               & \textbf{CIFAR-10}                                        & 86.42\%                  & 1.00                         & 86.31\%                     & 7.52e-50                        \\
                                                       &                                                               & \textbf{STL-10}                                          & 94.91\%                  & 1.00                         & 92.75\%                     & 9.22e-32                        \\
                                                       &                                                               & \textbf{SVHN}                                            & 71.60\%                  & 1.00                         & 70.71\%                     & 6.19e-70                        \\
                                                       &                                                               & \textbf{GTSRB}                                           & 85.54\%                  & 0.99                         & 84.60\%                     & 5.17e-52                        \\ \hline
\end{tabular}
}
\end{table}

\noindent\textbf{Negative Suspect Model.} These models should not be considered as unauthorized copies of the victim encoder, and thus, \name must not extract any watermark information. We argue that the training process of SSL encoders is intrinsically unique, influenced by factors such as the choice of optimization algorithm and the sequence in which training data is presented~\cite{dai2025division}. Crucially, these models are not derived from the victim encoder and should not retain the embedded watermark. We let the encoder architectures of these negative models be consistent with that of the victim encoder but differ in terms of hyperparameters and pretraining data configurations.  We design four variants: (1) Neg-v1, which differs only in pretraining data while keeping hyperparameters constant; (2) Neg-v2, which uses the same training data but differs in hyperparameters; (3) Neg-v3, which maintains both hyperparameters and pretraining data identical to the original model; and (4) Neg-v4, which differs in both hyperparameters and pretraining data. These negative models are developed independently and should not be considered stolen copies of the victim encoder, even when trained with the same dataset and hyperparameters as the victim encoder.

\subsection{Effectiveness}

\par Firstly, we primarily consider the most common DT attacker in MLaaS scenario, where the EaaS scenario is defered in Appendix~\ref{sec:appendix_eaas}: for a watermarked encoder, we adopt the standard transfer learning procedure outlined in BadEncoder \cite{jia2022badencoder}, which involves freezing the encoder and fine-tuning the classifier. The overall transfer process follows the established protocol described in \cite{lv2022ssl}. As shown in TABLE \ref{tab:four_ssl_performance}, we evaluate the ACC\textsubscript{\textit{m}} and $p$-values for ownership verification when transferring encoders pre-trained on CIFAR-10, Imagenette, and ImageNet, respectively, to downstream tasks under four different SSL algorithms. 

The experimental results demonstrate that the $p$-values of the watermarked downstream models (i.e., those stolen by the DT attacker) are significantly lower than the predefined threshold of $\tau=0.05$, with a maximum value of only 7.16e-17, thus successfully identifying them as pirated models. Meanwhile, \name applies to all four mainstream SSL algorithms, and both the embedding and verification of the watermark remain unaffected by the choice of SSL methods, demonstrating the \name's broad applicability.

\par The $p$-values of the clean downstream models in TABLE \ref{tab:four_ssl_performance} range between 0.98 and 1.00, significantly exceeding the threshold of 0.05, thus confirming their legitimacy as non-pirated models. To rule out effects of randomness, we construct 64 SimCLR-based negative suspect encoders by varying hyperparameters or training data, and transfer them to downstream tasks for large-scale ownership verification tests. TABLE \ref{tab:negative_positive_pvalue} presents the average verification results across four types of negative suspect models in downstream scenarios. The $p$-values remain within the range of 0.98-1.00, and the overall FPR is as low as 0.00\%. These results indicate that the presence of the watermark in the shadow probing pairs does not alter the prediction outputs of models without embedded watermarks, thereby demonstrating the fairness of the proposed ownership verification mechanism.

\subsection{Number of Probing Pairs}
We further examine the impact of probing pair quantity on \name's ownership verification in MLaaS, using the ResNet-18 encoder pretrained on Imagenette. As shown in TABLE~\ref{tab:nums_pairs}, we assess the $p$-values across different downstream classifiers using the numbers of 20, 50, 100, 200, and 500 probing pairs. The results indicate that \name can successfully verify ownership even with as few as 20 probing pairs. As the number of probes increases from 20 to 500, the $p$-values steadily decrease (further away from the threshold 0.05), suggesting that both statistical significance and verification reliability improve with more probes. Overall, we find that using 200 probing pairs provides an optimal balance between verification reliability and computational overhead.

\begin{table}[!t]
\centering
\caption{The $p$-value of negative and positive suspect downstream classifiers with \name watermark.}
\label{tab:negative_positive_pvalue}
\scalebox{0.6}{
\begin{tabular}{wc{1.5cm}|wc{1.9cm}|wc{0.6cm}|wc{0.6cm}|wc{0.6cm}|wc{0.6cm}|wc{1.0cm}|wc{1.0cm}|wc{1.0cm}|wc{1.0cm}}
\hline
\multirow{2}{*}{\textbf{\makecell[c]{Pre-training\\Task}}} & \multirow{2}{*}{\textbf{\makecell[c]{Downstream\\Task}}} & \multicolumn{4}{c|}{\textbf{Negative Suspect Model}} & \multicolumn{4}{c}{\textbf{Positive Suspect Model}} \\ \cline{3-10} 
                                                           &                                                           & \textbf{Neg-1} & \textbf{Neg-2} & \textbf{Neg-3} & \textbf{Neg-4} & \textbf{DT} & \textbf{FT} & \textbf{PR-15\%} & \textbf{PR-60\%} \\ \hline
\multirow{4}{*}{\textbf{CIFAR-10}}                                  & \textbf{CIFAR-10}                                                 & 1.00           & 1.00           & 1.00           & 1.00           & 2.18e-25    & 3.75e-22    & 1.92e-25    & 4.26e-17    \\ 
                                                           & \textbf{CINIC-10}                                                 & 1.00           & 1.00           & 1.00           & 1.00           & 3.16e-18    & 2.56e-18    & 7.43e-18    & 1.87e-16    \\ 
                                                           & \textbf{STL-10}                                                   & 1.00           & 1.00           & 1.00           & 1.00           & 6.08e-16    & 2.45e-10    & 3.31e-10    & 5.79e-12    \\ 
                                                           & \textbf{GTSRB}                                                    & 0.99           & 1.00           & 0.99           & 1.00           & 1.22e-22    & 6.83e-23    & 3.28e-21    & 7.14e-17    \\ \hline
\multirow{6}{*}{\textbf{ImageNet}}                                  & \textbf{Tiny-ImageNet}                                            & 1.00           & 1.00           & 1.00           & 1.00           & 8.26e-90    & 9.47e-29    & 2.94e-51    & 1.18e-47    \\ 
                                                           & \textbf{CIFAR-100}                                                & 1.00           & 1.00           & 1.00           & 1.00           & 3.37e-47    & 4.82e-65    & 1.39e-90    & 6.28e-55    \\ 
                                                           & \textbf{CIFAR-10}                                                 & 1.00           & 1.00           & 1.00           & 1.00           & 9.68e-34    & 2.14e-28    & 3.47e-51    & 1.92e-21    \\ 
                                                           & \textbf{STL-10}                                                   & 1.00           & 1.00           & 1.00           & 1.00           & 1.68e-84    & 2.38e-17    & 1.25e-20    & 3.96e-17    \\ 
                                                           & \textbf{SVHN}                                                     & 1.00           & 1.00           & 1.00           & 1.00           & 4.92e-42    & 3.57e-32    & 6.88e-64    & 1.32e-39    \\ 
                                                           & \textbf{GTSRB}                                                    & 0.98           & 0.98           & 0.99           & 1.00           & 8.26e-90    & 9.87e-44    & 2.74e-51    & 3.45e-30    \\ \hline
\end{tabular}}
\end{table}

\begin{table}[!t]
\centering
\caption{The $p$-value of \name's downstream classifiers under different numbers of probing pairs.} 
\label{tab:nums_pairs}
\scalebox{0.65}{
\begin{tabular}{c|cccc}
\hline
\multirow{2}{*}{\textbf{Number}} & \multicolumn{4}{c}{\textbf{Downstream Task}}                                                                                          \\ \cline{2-5} 
                                                  & \multicolumn{1}{c|}{\textbf{ImageNette}} & \multicolumn{1}{c|}{\textbf{CIFAR-10}} & \multicolumn{1}{c|}{\textbf{STL-10}} & \textbf{SVHN} \\ \hline
\textbf{20}                                       & \multicolumn{1}{c|}{1.19e-07}            & \multicolumn{1}{c|}{1.80e-2}           & \multicolumn{1}{c|}{2.85e-07}        & 1.06e-2       \\ \hline
\textbf{50}                                       & \multicolumn{1}{c|}{3.24e-19}            & \multicolumn{1}{c|}{1.00e-4}           & \multicolumn{1}{c|}{2.57e-14}        & 2.31e-06      \\ \hline
\textbf{100}                                      & \multicolumn{1}{c|}{1.92e-36}            & \multicolumn{1}{c|}{1.09e-08}          & \multicolumn{1}{c|}{1.68e-29}        & 1.51e-10      \\ \hline
\textbf{200}                                      & \multicolumn{1}{c|}{8.39e-72}            & \multicolumn{1}{c|}{4.52e-17}          & \multicolumn{1}{c|}{8.57e-58}        & 6.70e-18      \\ \hline
\textbf{500}                                      & \multicolumn{1}{c|}{2.08e-139}           & \multicolumn{1}{c|}{6.36e-27}          & \multicolumn{1}{c|}{1.53e-112}       & 1.39e-36      \\ \hline
\end{tabular}}
\end{table}

\subsection{Computation Overhead}

\begin{figure}[!t]
  \centering
    \includegraphics[width=0.5\textwidth]{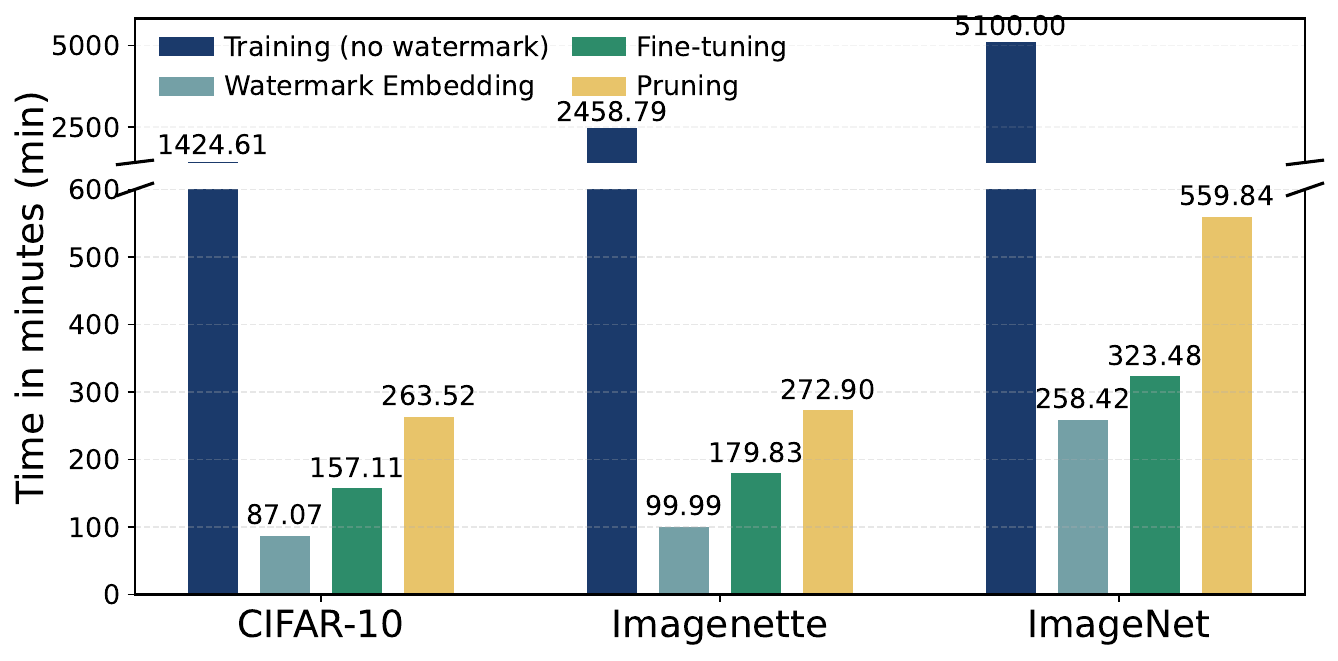}
  \caption{Overhead of \name's watermark embedding.}
  \label{fig:overhead}
\end{figure}

\par The computation overhead of \name under SimCLR mainly originates from two aspects: watermark embedding into the encoder by the model owner and ownership verification for suspect models. As illustrated in Fig. \ref{fig:overhead}, the overhead for watermark embedding accounts for only a minor proportion of the total training time across three pre-trained datasets: 6.11\% on CIFAR-10, 4.07\% on Imagenette, and 7.66\% on ImageNet. Meanwhile, the ownership verification process requires approximately 10 seconds to allow 200 probing pairs to query, making its computational cost practically negligible.

In terms of attack cost, given that potential model infringers generally operate with constrained computational resources, we evaluate the time required to execute model fine-tuning and pruning attacks. Experimental results indicate that both attacks demand substantially more time than watermark embedding. Specifically, on CIFAR-10 and Imagenette, the time cost of pruning attacks is approximately three times that of watermark embedding, whereas on ImageNet, it is about twice as much.
Overall, compared to the substantial computational resources required for model theft and encoder training, our watermark scheme provides reliable IP verification with minimal additional overhead, establishing its practical efficiency and advantages in real-world deployment scenarios.

\subsection{Robustness against Watermark Removal}
\par In addition to the previously natural and intuitive DT-type attackers, we further consider when the attacker is aware that a watermark has been embedded within the victim encoder and thus actively employs techniques such as model fine-tuning and pruning in an attempt to remove the watermark. 
It is noted that the attack typically knows the SSL pre-training task but does not have access to the original pre-training data. Therefore, in a real-world attack, the adversary first transfers the SSL encoder to the downstream task and then uses the downstream dataset to carry out the watermark removal.

\begin{table*}[t]
  \centering
  
  \begin{minipage}[t]{0.66\textwidth} 
    \centering
    \caption{Comparison of the ACC\textsubscript{\textit{m}} and $p$-value under fine-tuning and pruning settings.}
    \label{tab:ft_pruning_performance}
    \scalebox{0.58}{
    \begin{tabular}{c|cc|wc{1.5cm}|ccccccc}
    \hline
    \textbf{Pre-training}      & \multicolumn{2}{c|}{\textbf{Downstream}}                        & \multirow{2}{*}{\textbf{Fine-Tuning}} & \multicolumn{7}{c}{\textbf{Pruning Rate}} \\ \cline{5-11} 
    \textbf{Task}           & \multicolumn{2}{c|}{\textbf{Task}}                              &                                       & \multicolumn{1}{c|}{\textbf{5\%}} & \multicolumn{1}{c|}{\textbf{15\%}} & \multicolumn{1}{c|}{\textbf{40\%}} & \multicolumn{1}{c|}{\textbf{60\%}} & \multicolumn{1}{c|}{\textbf{75\%}} & \textbf{85\%}                 & \textbf{99\%} \\ \hline
    \multirow{8}{*}{\textbf{CIFAR-10}}  & \multicolumn{1}{c|}{\multirow{2}{*}{\textbf{CIFAR-10}}}      & ACC\textsubscript{\textit{m}}       & 86.31\%                               & \multicolumn{1}{c|}{86.60\%}     & \multicolumn{1}{c|}{85.40\%}       & \multicolumn{1}{c|}{84.10\%}       & \multicolumn{1}{c|}{83.60\%}       & \multicolumn{1}{c|}{83.20\%}       & \multicolumn{1}{c|}{81.30\%}  & 10.70\%       \\ \cline{3-11} 
                               & \multicolumn{1}{c|}{}                               & $p$-value & 4.21e-23                              & \multicolumn{1}{c|}{4.20e-26}     & \multicolumn{1}{c|}{2.36e-25}      & \multicolumn{1}{c|}{1.35e-22}      & \multicolumn{1}{c|}{5.10e-18}      & \multicolumn{1}{c|}{3.23e-15}      & \multicolumn{1}{c|}{3.46e-12} & 1.00          \\ \cline{2-11} 
                               & \multicolumn{1}{c|}{\multirow{2}{*}{\textbf{CINIC}}}         & ACC\textsubscript{\textit{m}}       & 74.03\%                               & \multicolumn{1}{c|}{73.30\%}      & \multicolumn{1}{c|}{72.30\%}       & \multicolumn{1}{c|}{71.20\%}       & \multicolumn{1}{c|}{70.30\%}       & \multicolumn{1}{c|}{69.20\%}       & \multicolumn{1}{c|}{66.30\%}  & 11.60\%       \\ \cline{3-11} 
                               & \multicolumn{1}{c|}{}                               & $p$-value & 1.42e-19                              & \multicolumn{1}{c|}{2.38e-18}     & \multicolumn{1}{c|}{8.82e-19}      & \multicolumn{1}{c|}{1.20e-18}      & \multicolumn{1}{c|}{2.18e-17}      & \multicolumn{1}{c|}{7.04e-15}      & \multicolumn{1}{c|}{8.14e-12} & 1.00          \\ \cline{2-11} 
                               & \multicolumn{1}{c|}{\multirow{2}{*}{\textbf{STL-10}}}        & ACC\textsubscript{\textit{m}}       & 77.00\%                               & \multicolumn{1}{c|}{77.80\%}      & \multicolumn{1}{c|}{77.90\%}       & \multicolumn{1}{c|}{76.50\%}       & \multicolumn{1}{c|}{75.60\%}       & \multicolumn{1}{c|}{72.02\%}       & \multicolumn{1}{c|}{68.00\%}  & 10.30\%       \\ \cline{3-11} 
                               & \multicolumn{1}{c|}{}                               & $p$-value & 1.77e-11                              & \multicolumn{1}{c|}{4.44e-11}     & \multicolumn{1}{c|}{1.62e-11}      & \multicolumn{1}{c|}{1.70e-14}      & \multicolumn{1}{c|}{6.88e-13}      & \multicolumn{1}{c|}{2.56e-10}      & \multicolumn{1}{c|}{2.56e-9}  & 1.00          \\ \cline{2-11} 
                               & \multicolumn{1}{c|}{\multirow{2}{*}{\textbf{GTSRB}}}         & ACC\textsubscript{\textit{m}}       & 78.10\%                               & \multicolumn{1}{c|}{76.20\%}      & \multicolumn{1}{c|}{75.70\%}       & \multicolumn{1}{c|}{74.00\%}       & \multicolumn{1}{c|}{72.10\%}       & \multicolumn{1}{c|}{70.70\%}       & \multicolumn{1}{c|}{65.60\%}  & 12.40\%       \\ \cline{3-11} 
                               & \multicolumn{1}{c|}{}                               & $p$-value & 5.21e-24                              & \multicolumn{1}{c|}{1.38e-22}     & \multicolumn{1}{c|}{4.10e-22}      & \multicolumn{1}{c|}{8.80e-17}      & \multicolumn{1}{c|}{2.76e-16}      & \multicolumn{1}{c|}{2.76e-13}      & \multicolumn{1}{c|}{0.0098}   & 1.00          \\ \hline
    \multirow{12}{*}{\textbf{ImageNet}} & \multicolumn{1}{c|}{\multirow{2}{*}{\textbf{Tiny-ImageNet}}} & ACC\textsubscript{\textit{m}}       & 75.92\%                               & \multicolumn{1}{c|}{72.00\%}      & \multicolumn{1}{c|}{71.92\%}       & \multicolumn{1}{c|}{70.21\%}       & \multicolumn{1}{c|}{69.77\%}       & \multicolumn{1}{c|}{68.98\%}       & \multicolumn{1}{c|}{65.43\%}  & 20.20\%       \\ \cline{3-11} 
                               & \multicolumn{1}{c|}{}                               & $p$-value & 4.12e-20                              & \multicolumn{1}{c|}{6.14e-55}     & \multicolumn{1}{c|}{3.78e-52}      & \multicolumn{1}{c|}{9.21e-50}      & \multicolumn{1}{c|}{1.45e-48}      & \multicolumn{1}{c|}{7.63e-45}      & \multicolumn{1}{c|}{2.09e-26} & 1.00          \\ \cline{2-11} 
                               & \multicolumn{1}{c|}{\multirow{2}{*}{\textbf{CIFAR-100}}}     & ACC\textsubscript{\textit{m}}       & 82.61\%                               & \multicolumn{1}{c|}{72.09\%}      & \multicolumn{1}{c|}{72.19\%}       & \multicolumn{1}{c|}{71.40\%}       & \multicolumn{1}{c|}{70.90\%}       & \multicolumn{1}{c|}{67.31\%}       & \multicolumn{1}{c|}{65.04\%}  & 15.42\%       \\ \cline{3-11} 
                               & \multicolumn{1}{c|}{}                               & $p$-value & 5.36e-66                              & \multicolumn{1}{c|}{6.66e-91}     & \multicolumn{1}{c|}{1.57e-91}      & \multicolumn{1}{c|}{3.28e-80}      & \multicolumn{1}{c|}{7.05e-56}      & \multicolumn{1}{c|}{6.66e-46}      & \multicolumn{1}{c|}{9.43e-40} & 1.00          \\ \cline{2-11} 
                               & \multicolumn{1}{c|}{\multirow{2}{*}{\textbf{CIFAR-10}}}      & ACC\textsubscript{\textit{m}}       & 94.84\%                               & \multicolumn{1}{c|}{89.85\%}      & \multicolumn{1}{c|}{89.65\%}       & \multicolumn{1}{c|}{88.88\%}       & \multicolumn{1}{c|}{82.46\%}       & \multicolumn{1}{c|}{80.19\%}       & \multicolumn{1}{c|}{78.16\%}  & 25.19\%       \\ \cline{3-11} 
                               & \multicolumn{1}{c|}{}                               & $p$-value & 2.52e-29                              & \multicolumn{1}{c|}{7.58e-52}     & \multicolumn{1}{c|}{4.12e-52}      & \multicolumn{1}{c|}{8.76e-46}      & \multicolumn{1}{c|}{2.35e-22}      & \multicolumn{1}{c|}{6.91e-26}      & \multicolumn{1}{c|}{1.57e-22} & 1.00          \\ \cline{2-11} 
                               & \multicolumn{1}{c|}{\multirow{2}{*}{\textbf{STL-10}}}        & ACC\textsubscript{\textit{m}}       & 96.04\%                               & \multicolumn{1}{c|}{95.00\%}      & \multicolumn{1}{c|}{94.13\%}       & \multicolumn{1}{c|}{93.07\%}       & \multicolumn{1}{c|}{90.80\%}       & \multicolumn{1}{c|}{84.36\%}       & \multicolumn{1}{c|}{82.38\%}  & 10.00\%       \\ \cline{3-11} 
                               & \multicolumn{1}{c|}{}                               & $p$-value & 2.61e-18                              & \multicolumn{1}{c|}{9.11e-21}     & \multicolumn{1}{c|}{1.46e-21}      & \multicolumn{1}{c|}{7.31e-20}      & \multicolumn{1}{c|}{4.87e-18}      & \multicolumn{1}{c|}{3.77e-16}      & \multicolumn{1}{c|}{1.01e-11} & 1.00          \\ \cline{2-11} 
                               & \multicolumn{1}{c|}{\multirow{2}{*}{\textbf{SVHN}}}          & ACC\textsubscript{\textit{m}}       & 95.57\%                               & \multicolumn{1}{c|}{66.00\%}      & \multicolumn{1}{c|}{65.30\%}       & \multicolumn{1}{c|}{64.41\%}       & \multicolumn{1}{c|}{61.93\%}       & \multicolumn{1}{c|}{60.40\%}       & \multicolumn{1}{c|}{58.59\%}  & 13.55\%       \\ \cline{3-11} 
                               & \multicolumn{1}{c|}{}                               & $p$-value & 4.15e-33                              & \multicolumn{1}{c|}{1.54e-70}     & \multicolumn{1}{c|}{8.02e-65}      & \multicolumn{1}{c|}{9.32e-65}      & \multicolumn{1}{c|}{1.54e-40}      & \multicolumn{1}{c|}{3.21e-35}      & \multicolumn{1}{c|}{6.90e-25} & 1.00          \\ \cline{2-11} 
                               & \multicolumn{1}{c|}{\multirow{2}{*}{\textbf{GTSRB}}}         & ACC\textsubscript{\textit{m}}       & 93.48\%                               & \multicolumn{1}{c|}{77.54\%}      & \multicolumn{1}{c|}{75.97\%}       & \multicolumn{1}{c|}{74.61\%}       & \multicolumn{1}{c|}{72.35\%}       & \multicolumn{1}{c|}{72.19\%}       & \multicolumn{1}{c|}{68.31\%}  & 10.00\%       \\ \cline{3-11} 
                               & \multicolumn{1}{c|}{}                               & $p$-value & 1.09e-44                              & \multicolumn{1}{c|}{2.98e-59}     & \multicolumn{1}{c|}{3.29e-52}      & \multicolumn{1}{c|}{2.46e-50}      & \multicolumn{1}{c|}{4.17e-31}      & \multicolumn{1}{c|}{1.51e-28}      & \multicolumn{1}{c|}{6.21e-9}  & 1.00          \\ \hline
    \end{tabular}}
  \end{minipage}
  \hfill 
  \begin{minipage}[t]{0.32\textwidth} 
    \centering
    \caption{Detection results by MM-BD~\cite{wang2024mm}.} 
    \label{tab:mmbd}
    \scalebox{0.6}{
    \begin{tabular}{c|p{0.2cm}p{0.2cm}p{0.2cm}p{0.2cm}p{0.6cm}|p{0.2cm}p{0.2cm}p{0.2cm}p{0.2cm}p{0.6cm}}
    \hline
    \multirow{2}{*}{\textbf{\begin{tabular}[c]{@{}c@{}}Downstream\\ Task\end{tabular}}} & \multicolumn{5}{c|}{\textbf{Ours}} & \multicolumn{5}{c}{\textbf{SSL-WM}} \\ \cline{2-11} 
    & \textbf{TP} & \textbf{FP} & \textbf{FN} & \textbf{TN} & \textbf{ACC\textsubscript{\textit{d}}} & \textbf{TP} & \textbf{FP} & \textbf{FN} & \textbf{TN} & \textbf{ACC\textsubscript{\textit{d}}} \\ \hline
    \textbf{CIFAR-10} & 0 & 0 & 64 & 64 & \textcolor{blue}{50\%} & 64 & 0 & 0 & 64 & \textcolor{red}{100\%} \\
    \textbf{CINIC-10} & 0 & 0 & 64 & 64 & \textcolor{blue}{50\%} & 64 & 0 & 0 & 64 & \textcolor{red}{100\%} \\
    \textbf{STL-10} & 1 & 0 & 63 & 64 & \textcolor{blue}{51\%} & 64 & 0 & 0 & 64 & \textcolor{red}{100\%} \\
    \textbf{ImgNet} & 4 & 0 & 64 & 64 & \textcolor{blue}{53\%} & 64 & 0 & 0 & 64 & \textcolor{red}{100\%} \\ \hline
    \end{tabular}}
    
    \vspace{0.8cm} 
    
    \caption{Detection results by DECREE~\cite{feng2023detecting}.} 
    \label{tab:DECREE}
    \scalebox{0.68}{ 
    \begin{tabular}{c|ccccc}
    \hline
    \multirow{2}{*}{\textbf{Methods}}      & \multicolumn{5}{c}{\textbf{Metrics}} \\ \cline{2-6} 
                                           & \textbf{TP} & \textbf{FP} & \textbf{FN} & \textbf{TN} & \textbf{ACC\textsubscript{\textit{d}}} \\ \hline
    \textbf{Ours}                          & 8           & 2           & 56          & 62           & \textcolor{blue}{54.69\%}              \\
    \textbf{SSL-WM}                        & 64          & 1           & 0           & 63           & \textcolor{red}{99.22\%}            \\
    \textbf{SSLGuard}                      & 64          & 1           & 0           & 63           & \textcolor{red}{99.22\%}            \\ \hline
    \end{tabular}}
  \end{minipage}
\end{table*}

\par \textbf{Robustness against fine-tuning.} In the downstream attack scenario permitting full fine-tuning of all layers, we conduct the FT attack on the watermarked model using stochastic gradient descent (SGD) with a high learning rate of $\eta = 1\times10^{-3}$ and a decay factor of $1\times10^{-6}$. The model is optimized until full convergence is achieved. As summarized in TABLE \ref{tab:ft_pruning_performance}, the IP verification $p$-value of the FT model on the shadow dataset shows a slight increase compared to that of the original DT model in TABLE \ref{tab:four_ssl_performance}. Nevertheless, it remains significantly below the predetermined threshold (i.e., 0.05), confirming the robustness of our watermark against the model FT attack.

\par \textbf{Robustness against pruning.} We conduct a series of pruning attack experiments using the SimCLR-based watermarked encoders. The watermarked encoders on CIFAR-10 and ImageNet are fine-tuned to the downstream task and subsequently pruned with pruning ratios ranging from 5\% to 99\%. As illustrated in TABLE \ref{tab:ft_pruning_performance}, the ACC\textsubscript{\textit{m}} of the watermarked model decreases progressively as the pruning ratio increases. Regarding ownership verification, when the pruning ratio is below 40\%, the $p$-value remains close to that of the unpruned DT model. At higher pruning ratios between 60\% and 85\%, although the $p$-value shows a moderate increase, it still remains significantly below the threshold of 0.05. When the pruning ratio reaches 99\%, the model utility severely deteriorates, with accuracies of ResNet-18 and ResNet-50 dropping to $10.00\% \sim 25.19\%$, rendering ownership verification practically irrelevant in such scenarios. The observed robustness is attributed to the phenomenon reported in work \cite{lv2022ssl}: watermark samples and task-specific samples tend to activate overlapping sets of neurons. Therefore, the proposed watermark remains verifiable as long as the model maintains its utility for the downstream task.

\par We construct 64 positive suspect encoders through DT, FT, and two pruning configurations (PR-$15$\% and PR-$60$\%), and adapt them to downstream tasks for large-scale ownership verification tests. As shown in TABLE \ref{tab:negative_positive_pvalue}, the average $p$-values obtained from the four types of positive suspect models across different downstream tasks remain significantly below the decision threshold of 0.05. The batch-experimental results further confirm a watermark detection rate of 100\%, successfully identifying all suspect models as derived from the watermarked encoder.

\subsection{Robustness against Watermark Detection}
\par After stealing target encoders, attackers may detect whether the watermark exists. Specifically, the attacker could leverage MM-BD \cite{wang2024mm} for the downstream classifier's output analysis or employ DECREE \cite{feng2023detecting} for trigger inversion. The watermark detection accuracy (ACC\textsubscript{\textit{d}}) measures the proportion of models in which the watermark is correctly detected by an attacker.

\par \textbf{MM-BD \cite{wang2024mm}.} The offline MM-BD identifies watermarked models by analyzing the backdoor effects introduced by watermarks in classifier outputs, computing a maximum gap statistic for each class to measure the discrepancy between the watermarked target class and other classes. We apply the MM-BD detector to evaluate 64 watermarked and 64 clean downstream models. The results in TABLE \ref{tab:mmbd} show that all clean models are correctly identified as negative, with an FPR of 0.00\%, indicating that MM-BD maintains high specificity under normal conditions. For the 64 \name watermarked models, MM-BD achieves an ACC\textsubscript{\textit{d}} of 50.00\%-53.00\%, which is close to random guessing. This demonstrates that our watermarking scheme effectively evades detection by the MM-BD framework.
We attribute this evasion capability to our watermark design strategy, which promotes high dispersion of watermarked sample representations in the feature space while minimizing their pairwise cosine similarity. This approach prevents watermarked samples from being classified into the same class in the downstream prediction space, thereby avoiding the creation of a detectable backdoor effect. Consequently, MM-BD, which relies on identifying such statistical anomalies, fails to detect the presence of \name
.

\par \textbf{DECREE \cite{feng2023detecting}.} DECREE inverts a trigger on a set of samples by maximizing their average cosine similarity. It determines that an encoder is backdoored if the inverted is smaller than a certain threshold (10\% of the input image size in the original paper). We evaluate DECREE on 64 clean and 64 watermarked encoders pre-trained on CIFAR-10 and Imagenette for each watermarking method. As shown in TABLE \ref{tab:DECREE}, DECREE achieves a 3.23\% FPR on clean encoders and successfully detects watermarks in both SSLGuard and SSL-WM with ACC\textsubscript{\textit{d}} = 99\%. In contrast, when applied to our method, it achieves the ACC\textsubscript{\textit{d}} = 54.69\%. This result can be attributed to our watermark design: watermarked samples are dispersed across non-source classes, which substantially reduces their feature density and average cosine similarity, thereby effectively evading similarity-based trigger inversion.

\subsection{Utility}
\par We evaluate the ACC\textsubscript{\textit{m}} of multiple models under different configurations, including: clean models (i.e., models without embedded watermarks), watermarked models, fully fine-tuned models on downstream datasets, and pruned downstream models. As shown in TABLE \ref{tab:four_ssl_performance}, the accuracy comparison between clean and watermarked models demonstrates that the embedding of watermarks has a negligible overall impact on model performance. Specifically, for the ResNet-18 encoder pre-trained on CIFAR-10, the accuracy degradation of four watermarked SSL encoders on downstream tasks is as follows: 0.25\% for SimCLR, 0.61\% for MoCo v2, 0.43\% for BYOL, and 0.60\% for SimSiam. In contrast, the ResNet-50 encoder pre-trained on ImageNet exhibits greater sensitivity to watermarks, with an average accuracy drop of 1.13\% on downstream tasks. We hypothesize the reasons as follows: due to the lower resolution and relatively simpler scenes in the CIFAR-10 dataset, the feature dimensions that the model needs to learn are limited. The parameter perturbations and feature space variations introduced by the watermark can be absorbed by the relatively sufficient model capacity, thus resulting in a minimal impact on final classification performance. In comparison, the ImageNet dataset is larger in scale, exhibits finer-grained categories, and possesses more complex and high-dimensional feature structures. The modifications to weight parameters and alterations in feature distributions during watermark embedding are more likely to cause cumulative interference in highly refined representation learning, thereby leading to a more noticeable degradation in transfer performance on downstream tasks.
\par Furthermore, as presented in TABLE \ref{tab:ft_pruning_performance}, when the encoder parameters are fine-tuned on downstream tasks for 30-100 epochs until full convergence, the watermark models not only maintain reliable ownership verification but also exhibit improved accuracy. Specifically, the watermarked ResNet-18 model pre-trained on CIFAR-10 slightly outperforms its clean counterpart, while the watermarked ResNet-50 model pre-trained on ImageNet achieves significantly higher accuracy than the clean model. To evaluate robustness against model compression, we perform structured pruning with reducing parameter counts by 5\%, 15\%, 40\%, 60\%, 75\%, 85\%, and 99\%. The results demonstrate that the watermarked models retain high accuracy even after removing up to 85\% of the parameters. Only when the pruning ratio reaches 99\%, the ACC\textsubscript{\textit{m}} drops to an unacceptable level, which ranges between 10.30\% to 12.40\% for ResNet-18 and 10.00\% to 25.19\% for ResNet-50.

\begin{table}[!t]
	\centering
	\caption{Comparison of the original watermarked models vs. the watermarked models after Overwriting and Unlearning.}
	
    \centering
    (a) Pretraining on CIFAR-10
    \vspace{0.2cm}
    \scalebox{0.65}{
    \begin{tabular}{c|c|c|c|c}
    \hline
    \multirow{2}{*}{\textbf{\begin{tabular}[c]{@{}c@{}}Downstream\\ Task\end{tabular}}} & \multirow{2}{*}{\textbf{Metrics}} & \multirow{2}{*}{\textbf{Before}} & \multirow{2}{*}{\textbf{\begin{tabular}[c]{@{}c@{}}After\\ Overwriting\end{tabular}}} & \multirow{2}{*}{\textbf{\begin{tabular}[c]{@{}c@{}}After\\ Unlearning\end{tabular}}} \\
     & & & & \\ \hline
    \multirow{2}{*}{\textbf{CIFAR-10}} & ACC\textsubscript{\textit{m}} & 86.61\% & 79.02\% & 84.29\% \\ \cline{2-5} 
     & $p$-value & 3.26e-45 & 4.86e-43 & 5.87e-21 \\ \hline
    \multirow{2}{*}{\textbf{CINIC}} & ACC\textsubscript{\textit{m}} & 73.23\% & 72.19\% & 72.46\% \\ \cline{2-5} 
     & $p$-value & 2.84e-19 & 7.00e-24 & 8.29e-26 \\ \hline
    \multirow{2}{*}{\textbf{STL-10}} & ACC\textsubscript{\textit{m}} & 77.72\% & 73.77\% & 76.28\% \\ \cline{2-5} 
     & $p$-value & 7.16e-17 & 3.07e-29 & 2.34e-19 \\ \hline
    \multirow{2}{*}{\textbf{GTSRB}} & ACC\textsubscript{\textit{m}} & 76.44\% & 68.81\% & 69.13\% \\ \cline{2-5} 
     & $p$-value & 1.59e-23 & 6.29e-39 & 1.68e-10 \\ \hline
    \end{tabular}}
	
    \centering
    (b) Pretraining on ImageNette
    \vspace{0.2cm}
    \scalebox{0.65}{
    \begin{tabular}{c|c|c|c|c}
    \hline
    \multirow{2}{*}{\textbf{\begin{tabular}[c]{@{}c@{}}Downstream\\ Task\end{tabular}}} & \multirow{2}{*}{\textbf{Metrics}} & \multirow{2}{*}{\textbf{Before}} & \multirow{2}{*}{\textbf{\begin{tabular}[c]{@{}c@{}}After\\ Overwriting\end{tabular}}} & \multirow{2}{*}{\textbf{\begin{tabular}[c]{@{}c@{}}After\\ Unlearning\end{tabular}}} \\
     & & & & \\ \hline
    \multirow{2}{*}{\textbf{ImageNette}} & ACC\textsubscript{\textit{m}} & 84.08\% & 75.18\% & 83.74\% \\ \cline{2-5} 
     & $p$-value & 3.96e-44 & 3.17e-56 & 2.54e-24 \\ \hline
    \multirow{2}{*}{\textbf{CIFAR-10}} & ACC\textsubscript{\textit{m}} & 69.80\% & 55.40\% & 65.80\% \\ \cline{2-5} 
     & $p$-value & 3.98e-30 & 1.23e-64 & 6.00e-29 \\ \hline
    \multirow{2}{*}{\textbf{STL-10}} & ACC\textsubscript{\textit{m}} & 58.60\% & 58.30\% & 56.90\% \\ \cline{2-5} 
     & $p$-value & 6.72e-90 & 1.01e-93 & 4.47e-45 \\ \hline
    \multirow{2}{*}{\textbf{SVHN}} & ACC\textsubscript{\textit{m}} & 63.93\% & 54.20\% & 61.29\% \\ \cline{2-5} 
     & $p$-value & 2.42e-20 & 9.24e-27 & 6.90e-13 \\ \hline
    \end{tabular}}
	
	\label{tab:overwriting_and_unlearning}
\end{table}

\section{Adaptive Attacks}\label{6}
We now consider three adaptive attacks, categorized based on attacker knowledge of \name: 1) Knowledgeable Attacker (I): the attacker knows the trigger pattern but is unaware of the embedding procedure and watermark samples; 2) Knowledgeable Attacker (II): the attacker has knowledge of the embedding procedure and watermark samples but does not possess the trigger; and 3) Knowledgeable Attacker (III): as the strongest threat model, the attacker has full knowledge, including the watermark samples, trigger, and complete embedding pipeline. The attacker even has access to the partial pre-training dataset to serve as an auxiliary dataset.

\noindent{$\bullet$} \textbf{Knowledgeable Attacker (I):} The attacker can perform the watermark overwriting attack based on the known trigger by embedding a new pseudo-watermark into the encoder to overwrite the original watermark. Correspondingly, from the attacker's perspective, we introduce the BadEncoder \cite{jia2022badencoder} as the specific scheme to embed this pseudo-watermark.

\par This experiment is conducted on encoders pre-trained on the CIFAR-10 and Imagenette. To simulate an overwriting attack using the BadEncoder framework, we inject our trigger into samples of class 8, causing them to be misclassified as class 9, while ensuring a backdoor attack success rate of over 80\%. As shown in TABLE \ref{tab:overwriting_and_unlearning}, compared to the original ownership verification results, the $p$-value obtained from the models after overwriting is lower, indicating a more statistically significant verification outcome. We attribute this phenomenon to the fact that the overwriting attack biases the classification of triggered samples toward class 9, thereby increasing the confidence of their assignment to other classes and indirectly reducing the confidence in our original class (class 0).

\begin{figure*}[!t]
  \centering
    \includegraphics[width=0.91\textwidth]{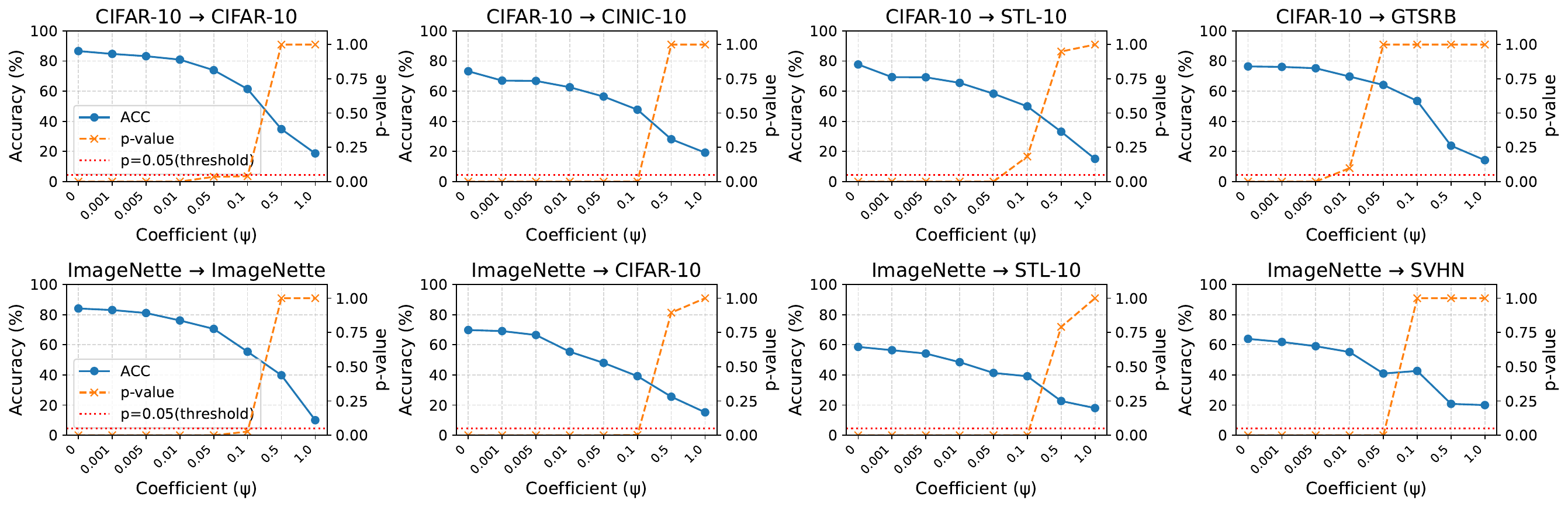}
  \caption{Knowledgeable Attacker (III) with knowing the watermark samples, trigger, and complete embedding pipeline.}
  \label{fig:adaptive_k3}
\end{figure*}


\noindent{$\bullet$} \textbf{Knowledgeable Attacker (II):} The attacker can leverage machine unlearning techniques \cite{li2025machine} to weaken or even erase the original watermark. Specifically, it can construct a new shadow dataset by injecting a chosen trigger into the clean samples in the shadow dataset $\mathcal{D}_{s}$ and then maximizing the similarity between the representations of clean samples and new triggered samples output by the watermarked encoder. TABLE \ref{tab:overwriting_and_unlearning} also presents the experimental results of machine unlearning attacks. The results show that although the $p$-value for ownership verification slightly increases in the watermarked models after unlearning, \name remains extractable, and ownership verification can still be achieved with a low $p$-value. This phenomenon demonstrates that our method can effectively resist such adaptive attacks.

\noindent{$\bullet$} \textbf{Knowledgeable Attacker (III):} To thoroughly circumvent potential IP disputes, such attackers may actively attempt to remove the watermark embedded in our encoder. This attacker can fine-tune the model by maximizing the loss function associated with the watermark term to eliminate the embedded watermark, as expressed in Eq. \ref{eq.6}.
\begin{equation}\label{eq.6}
\mathcal{L} = - \psi \min \frac{1}{S}\sum\limits_{i = 1}^S \left|\cos (\mathbf{e}_{\mathit{wm}}(x_{c,i}), \mathbf{e}_{\mathit{wm}}(x_{wm,i}))\right|.
\end{equation}

\par As illustrated in Fig. \ref{fig:adaptive_k3}, we measure the ACC\textsubscript{\textit{m}} and $p$-value under different values of $\psi$. Observations on encoders pre-trained on CIFAR-10 and Imagenette reveal that when $\psi$ = 0.5, the model's ACC\textsubscript{\textit{m}} drops sharply to an unusable level, leading to the failure of watermark verification. When $\psi$ = 0.1, on the downstream datasets (i.e., GTSRB, SVHN) with a significant distribution difference from the pre-training dataset, the attacker successfully removes our watermark at the cost of a 20\% decrease in ACC\textsubscript{\textit{m}}. Meanwhile, on other downstream datasets, although watermarks can still be extracted, their $p$-values are already very close to the threshold of 0.05. Under conditions where the model remains usable ($\psi <=$ 0.1), \name is still extractable and maintains a low $p$-value. Overall, effectively removing \name watermarks comes at the cost of significant model performance degradation, which contradicts the attacker's goal of profiting from stealing high-performance models. Moreover, this scenario assumes the attacker has full knowledge of all watermark-related details, which is an extremely improbable situation, further demonstrating the high adversarial robustness of \name.

\section{Conclusion}
In this work, we present \name, a unified and adversarial robust watermarking framework for pretrained SSL encoders. By incorporating an orthogonality-based trigger effect and feature-space constraints guided by latent representation entanglement and distribution alignment, \name effectively eliminates the detectability of watermarks while ensuring reliable IP black-box verification in both EaaS and MLaaS scenarios. Experimental results confirm that \name performs superior in terms of verification reliability, computation overhead, and robustness to watermark removal and adaptive attacks, while existing SOTAs cannot survive.



\begin{appendices}

\section{Algorithm}\label{sec:appendi_algo}

\subsection{Algorithm For Watermark Embedding}\label{sec:appendi_algo_1}

The algorithm for the watermark embedding is presented in Algorithm~\ref{algo:watermark embedding}.

\begin{algorithm}
    \caption{Watermark Embedding}
    \label{algo:watermark embedding}
    \begin{algorithmic}[1]
        \STATE \textbf{Input: The original pre-trained encoder $\mathbf{e}_{\mathit{c}}$, pre-training dataset $\mathcal{D} = \sum\limits_{i = 1}^N {x_i}$, shadow dataset $\mathcal{D}_{s} = \sum\limits_{i = 1}^S {({x_{c,i}},{x_{wm,i}})}$, anchor classes $A$, number of fine-tuing epochs $epoch$, warm epoch $E$, learning rate $\eta$, hyperparameter $\alpha$ and $\beta$.}
        \STATE \textbf{Output: The watermarked encoder $\mathbf{e}_{\mathit{wm}}$.}
        \STATE $\mathbf{e}_{\mathit{wm}} \leftarrow Copy(\mathbf{e}_{\mathit{c}})$
        \STATE $\mathbf{a}_{\textit{anchor}} \leftarrow ComputeClassCenters(\mathbf{e}_{\mathit{c}}, \mathcal{D}, A)$
        \FOR{each $e \in {1,2,...,epoch}$}
        \FOR{each batch $x$ in $\mathcal{D}$ and $(x_{c},x_{wm})$ in $\mathcal{D}_{s}$}
        \STATE $r_c^x \leftarrow ExtractRep(\mathbf{e}_{\mathit{c}},x)$
        \STATE $r_{wm}^x \leftarrow ExtractRep(\mathbf{e}_{\mathit{wm}},x)$
        \STATE $r_c \leftarrow ExtractRep(\mathbf{e}_{\mathit{wm}},x_{c})$
        \STATE $r_{wm} \leftarrow ExtractRep(\mathbf{e}_{\mathit{wm}},x_{wm})$
        \STATE $\mathcal{L}_{\mathit{ref}} \leftarrow MaxCosSim(r_c^x,r_{wm}^x)$ \hfill \textcolor{blue}{$\triangleright$ Eq. \ref{eq.2}}
        \STATE $\mathcal{L}_{\mathit{wm}} \leftarrow MinimizeCosSim(r_c,r_{wm})$ \hfill \textcolor{blue}{$\triangleright$ Eq. \ref{eq.3}}
        \STATE $\mathcal{L}_{\mathit{entgl}} \leftarrow MaxCosSim(r_{wm},\mathbf{a}_{\textit{anchor}})$ \hfill \textcolor{blue}{$\triangleright$ Eq. \ref{eq.4}}
        \STATE $\mathcal{L}_{\mathit{dis}} \leftarrow MinimizeSWD(r_{wm}^x,r_{wm})$ \hfill \textcolor{blue}{$\triangleright$ Eq. \ref{eq.5}}
        \IF{$epoch <= E$}
        \STATE $\mathcal{L}_{\mathit{total}} = \mathcal{L}_\mathit{ref} + \alpha \mathcal{L}_\mathit{wm}$
        \ELSE
        \STATE $\mathcal{L}_{\mathit{rbt}} = \mathcal{L}_\mathit{entgl}+\mathcal{L}_\mathit{dis}$
        \STATE $\mathcal{L}_{\mathit{total}} = \mathcal{L}_\mathit{ref} + \alpha \mathcal{L}_\mathit{wm} + \beta \mathcal{L}_\mathit{rbt}$
        \hfill \textcolor{blue}{$\triangleright$ Eq. \ref{eq.1}}
        \ENDIF
        \STATE $\mathbf{e}_{\mathit{wm}} \leftarrow \mathbf{e}_{\mathit{wm}} - \eta \nabla \mathcal{L}_{\mathit{total}}$
        \ENDFOR
        \ENDFOR
        \STATE \textbf{return} $\mathbf{e}_{\mathit{wm}}$
    \end{algorithmic}
\end{algorithm}

\subsection{Algorithm For Ownership Verification Under MLaaS Scenario}\label{sec:appendi_algo_2}

The algorithm for the ownership verification under MLaaS is presented in Algorithm~\ref{algo:ownership verification}.

\begin{algorithm}
    \caption{Ownership Verification Under MLaaS Scenario}
    \label{algo:ownership verification}
    \begin{algorithmic}[1]
        \STATE \textbf{Input: The suspect model $\tilde{\mathbf{f}}$, the verification dataset $\mathcal{D}_{s} = \sum\limits_{i = 1}^S {({x_{c,i}},{x_{wm,i}})}$, significant level $\lambda$.}
        \STATE \textbf{Output: The verification result: $True$ or $False$.}
        \FOR{each $({x_{c,i}},{x_{wm,i}})$ in $\mathcal{D}_{s}$}
        \STATE $\mathbf{P}_{c,i} \leftarrow \tilde{\mathbf{f}}(x_{c,i})$
        \STATE $\mathbf{P}_{wm,i} \leftarrow \tilde{\mathbf{f}}(x_{wm})$
        \ENDFOR
        \STATE $p$-value = $t$-test($\mathbf{P}_c$, $\mathbf{P}_{wm}$)  
        \IF{$p<=\lambda$}
        \STATE \textbf{return} $True$
        \ELSE
        \STATE \textbf{return} $False$
        \ENDIF
    \end{algorithmic}
\end{algorithm}

\section{Ownership Verification}\label{sec:appendi_Eaas}

\subsection{Ownership Verification Under Eaas Scenario }\label{sec:appendi_Eaas_veri}
In Eaas scenario, where the stolen encoder is directly deployed as a new EaaS, the model owner can query the suspect encoder $\tilde{\mathbf{e}}$ to obtain feature representations but has no access to internal parameters. To verify ownership, we evaluate the cosine similarity between probing pairs in $\mathcal{D}_s$ and formulate the verification task as a hypothesis test as described below:

\begin{proposition}
Let $\tilde{\mathbf{r}}_c = \tilde{\mathbf{e}}(x_c)$ and $\tilde{\mathbf{r}}_{wm} = \tilde{\mathbf{e}}(x_{wm})$ denote representations outputted by the suspect encoder $\tilde{\mathbf{e}}$ on probing pairs $(x_c, x_{wm})$, respectively. For paired samples, define the cosine similarity: $\tilde{M}_{c,{wm}} = |\cos(\tilde{\mathbf{e}}(x_{c}), \tilde{\mathbf{e}}(x_{wm}))|$. Let $\mu \in (0,1)$ denote the similarity threshold. The negative encoder, which lacks our \name watermark, will produce nearly identical representations for the paired clean and watermarked samples $(x_c, x_{wm})$, meaning the ideal value of $\tilde{M}_{c,{wm}}$ is 1. Given the null hypothesis $\mathbb{H}_0: 1 - \tilde{M}_{c,{wm}} \le \mu$ and the alternative hypothesis $\mathbb{H}_1: 1 - \tilde{M}_{c,{wm}} > \mu$, where the hyper-parameter $\mu \in (0,1)$, we claim that the suspect encoder is illegally derived from our well-trained encoder $\mathbf{e}_{\mathit{wm}}$ if and only if $\mathbb{H}_0$ is rejected.
\end{proposition}

If the $t$-test yields a $p$-value $p \le \lambda$ (significance level $\lambda=0.05$), we reject $H_0$ and claim that $\tilde{\mathbf{e}}$ originates from the our watermarked encoder. The algorithm for the ownership verification procedure in EaaS scenario is presented in Algorithm~\ref{algo:ownership verification I}.
\begin{algorithm}[t]
    \caption{Ownership Verification Under EaaS scenario}
    \label{algo:ownership verification I}
    \begin{algorithmic}[1]
        \STATE \textbf{Input: The suspect encoder $\tilde{\mathbf{e}}$, the verification dataset $\mathcal{D}_{s} = \sum_{i = 1}^S {({x_{c,i}},{x_{wm,i}})}$, significant level $\lambda$.}
        \STATE \textbf{Output: The verification result: $True$ or $False$.}
        \FOR{each $({x_{c,i}},{x_{wm,i}})$ in $\mathcal{D}_{s}$}
        \STATE $\tilde{\mathbf{r}}_{c,i} \leftarrow \tilde{\mathbf{e}}(x_{c,i})$
        \STATE $\tilde{\mathbf{r}}_{wm,i} \leftarrow \tilde{\mathbf{e}}(x_{wm})$
        \ENDFOR
        \STATE $p$-value = $t$-test(1, $\tilde{M}_{c,{wm}}$)  
        \IF{$p<=\lambda$}
        \STATE \textbf{return} $True$
        \ELSE
        \STATE \textbf{return} $False$
        \ENDIF
    \end{algorithmic}
\end{algorithm}

\section{Experiments}

\subsection{Details of Experimental Settings}
\label{appendix:experimental-settings}

\mypara{Datasets.} We use nine popular datasets for CV tasks (i.e., ImageNet, Tiny-ImageNet, Imagenette, CIFAR-100, CIFAR-10, STL-10, CINIC-10, SVHN, and GTSRB).

\noindent{$\bullet$} \textbf{ImageNet}~\cite{ImageNet} comprises 1.2 million training images categorized into 1,000 distinct classes. Each image has a resolution of $224 \times 224$ pixels with three color channels (RGB).

\noindent{$\bullet$} \textbf{Tiny-ImageNet}~\cite{tiny-imagenet} is a subset of the ImageNet dataset. It consists of 100,000 images spanning 200 classes, with each class containing 500 images for training and 50 for validation.

\noindent{$\bullet$} \textbf{Imagenette}~\cite{imagenette} is a subset of 10 classes from the ImageNet dataset. It has 9469 training images and 3925 test images. Each image size is $224 \times 224 \times 3$.

\noindent{$\bullet$} \textbf{CIFAR-100}~\cite{cifar100} contains 60, 000 images with size 32 × 32 × 3 in 100 classes. Each class contains 500 images for training and 100 images for testing.

\noindent{$\bullet$} \textbf{CIFAR-10}~\cite{cifar10} comprises 60,000 images with the size of 32 × 32 × 3, distributed across 10 distinct classes.

\noindent{$\bullet$} \textbf{STL-10}~\cite{coates2011analysis} contains 5,000 labeled images for training and 8,000 for testing, along with 100,000 unlabeled images, all conforming to the identical 10-class categorical structure defined by CIFAR-10.

\noindent{$\bullet$} \textbf{CINIC-10}~\cite{darlow2018cinic} is synthetically constructed by combining elements from ImageNet and CIFAR-10, comprising a total of 270,000 images. It retains the same 10-class taxonomic structure as CIFAR-10.

\noindent{$\bullet$} \textbf{SVHN}~\cite{svhn} is a real-world image dataset of 600,000+ digits obtained from house numbers in Google Street View images. The dataset consists of 10 classes, representing digits 0 to 9.

\noindent{$\bullet$} \textbf{GTSRB}~\cite{stallkamp2011german} is a dataset designed for traffic sign recognition, containing over 50,000 images of traffic signs categorized into 43 classes.

\mypara{Shadow Dataset.} Due to the inaccessibility of downstream task domains and their specific samples, a subset of samples is selected from the pre-training dataset to construct the shadow dataset for subsequent watermark embedding and verification. Specifically, when the pre-training dataset is CIFAR-10 and Imagenette, 200 samples are selected; for ImageNet, 500 samples are chosen. By embedding a trigger pattern into the selected samples, probing pairs consisting of clean samples and their watermarked samples is created.

\mypara{Platform.} All experiments are conducted on two separate hardware configurations: a primary workstation equipped with an NVIDIA GeForce RTX 3090 GPU (24 GB VRAM), Intel(R) Core(TM) i9-10900K CPU @ 3.70 GHz, and 32 GB RAM, and an auxiliary server featuring an NVIDIA GeForce RTX 4090 GPU (24 GB VRAM), 16 vCPU Intel(R) Xeon(R) Gold 6430 processor, and 120 GB RAM.

\subsection{Experimental Results of Ownership Verification Under EaaS Scenario}\label{sec:appendix_eaas}

\mypara{Effectiveness and Utility.} We first simulate the most common DT attacker under the EaaS scenario. As shown in TABLE~\ref{tab:eaas_encoder}, we take the SimCLR algorithm as an example to evaluate the classification accuracy (ACC\textsubscript{\textit{m}}) calculated by the $K$-Nearest Neighbors (KNN) algorithm and ownership verification $p$-values of the clean encoder, the \name watermarked encoder, and four negative encoders that are trained on the CIFAR-10, Imagenette, and ImageNet datasets, respectively.

The experimental results demonstrate that the FPR on negative models defined in Subsection~\ref{sec:5.1} is 0.00\%, confirming the reliability of our verification method proposed in Subsection~\ref{sec:appendi_Eaas_veri}. For the \name watermarked encoder, the $p$-values are all significantly lower than the predefined threshold $\tau=0.05$, with the maximum value of only 1.10e-39, successfully identifying these pirated encoders. Moreover, regarding encoder utility, \name introduces negligible degradation. Compared with the clean encoder, the \name's ACC\textsubscript{\textit{m}} decreases by only 1.28\% on CIFAR-10, 2.19\% on Imagenette, and 0.25\% on ImageNet, indicating that the watermarking process has minimal impact on encoder usability.

\begin{table*}[htbp]
\centering
\caption{Comparison of the ACC\textsubscript{\textit{m}} and $p$-value under clean and watermarked models pretrained by SimCLR.}
\label{tab:eaas_encoder} 
\scalebox{0.85}{
\begin{tabular}{c|wc{0.45cm}wc{0.7cm}|wc{0.45cm}wc{0.7cm}|wc{0.45cm}wc{0.7cm}|wc{0.45cm}wc{0.7cm}|wc{0.45cm}wc{0.7cm}|wc{0.45cm}wc{0.7cm}}
\hline
\multirow{2}{*}{\textbf{\begin{tabular}[c]{@{}c@{}} Dataset\end{tabular}}} & \multicolumn{2}{c|}{\textbf{Clean encoder}}            & \multicolumn{2}{c|}{\textbf{\name encoder}}            & \multicolumn{2}{c|}{\textbf{Neg-1}}                    & \multicolumn{2}{c|}{\textbf{Neg-2}}                    & \multicolumn{2}{c|}{\textbf{Neg-3}}                    & \multicolumn{2}{c}{\textbf{Neg-4}}                     \\ \cline{2-13} 
                                                                                      & \multicolumn{1}{c|}{\textbf{ACC\textsubscript{\textit{m}}}} & \textbf{$p$-value} & \multicolumn{1}{c|}{\textbf{ACC\textsubscript{\textit{m}}}} & \textbf{$p$-value} & \multicolumn{1}{c|}{\textbf{ACC\textsubscript{\textit{m}}}} & \textbf{$p$-value} & \multicolumn{1}{c|}{\textbf{ACC\textsubscript{\textit{m}}}} & \textbf{$p$-value} & \multicolumn{1}{c|}{\textbf{ACC\textsubscript{\textit{m}}}} & \textbf{$p$-value} & \multicolumn{1}{c|}{\textbf{ACC\textsubscript{\textit{m}}}} & \textbf{$p$-value} \\ \hline
\textbf{CIFAR-10}                                                                      & \multicolumn{1}{c|}{87.48\%}       & 1.00               & \multicolumn{1}{c|}{86.20\%}       & 2.98e-58           & \multicolumn{1}{c|}{79.34\%}       & 1.00               & \multicolumn{1}{c|}{89.41\%}       & 1.00               & \multicolumn{1}{c|}{85.15\%}       & 1.00               & \multicolumn{1}{c|}{80.14\%}       & 1.00               \\ \hline
\textbf{Imagenette}                                                                   & \multicolumn{1}{c|}{82.55\%}       & 1.00               & \multicolumn{1}{c|}{80.36\%}       & 2.82e-56           & \multicolumn{1}{c|}{82.91\%}       & 1.00               & \multicolumn{1}{c|}{67.08\%}       & 1.00               & \multicolumn{1}{c|}{64.8\%}        & 1.00               & \multicolumn{1}{c|}{83.13\%}       & 1.00               \\ \hline
\textbf{ImageNet}                                                                     & \multicolumn{1}{c|}{58.05\%}       & 0.99               & \multicolumn{1}{c|}{57.80\%}       & 1.10e-39           & \multicolumn{1}{c|}{62.10\%}       & 1.00               & \multicolumn{1}{c|}{59.04\%}       & 0.95               & \multicolumn{1}{c|}{48.60\%}       & 0.98               & \multicolumn{1}{c|}{62.45\%}       & 1.00               \\ \hline
\end{tabular}}
\end{table*}

\begin{table*}[htbp]
\centering
\caption{Comparison of the ACC\textsubscript{\textit{m}} and $p$-value under EaaS scenario after fine-tuning and pruning attacks.}
\label{tab:eaas_robustness_pruning} 
\scalebox{0.73}{
\begin{tabular}
{c|cc|cc|cc|cc|cc|cc|cc}
\hline
\multirow{2}{*}{\textbf{Dataset}} & \multicolumn{2}{c|}{\textbf{FT}}                       & \multicolumn{2}{c|}{\textbf{PR-5\%}}                    & \multicolumn{2}{c|}{\textbf{PR-20\%}}                   & \multicolumn{2}{c|}{\textbf{PR-40\%}}                   & \multicolumn{2}{c|}{\textbf{PR-60\%}}                   & \multicolumn{2}{c|}{\textbf{PR-85\%}}                   & \multicolumn{2}{c}{\textbf{PR-95\%}}                                         \\ \cline{2-15} 
                                  & \multicolumn{1}{c|}{\textbf{ACC\textsubscript{\textit{m}}}} & \textbf{$p$-value} & \multicolumn{1}{c|}{\textbf{ACC\textsubscript{\textit{m}}}} & \textbf{$p$-value} & \multicolumn{1}{c|}{\textbf{ACC\textsubscript{\textit{m}}}} & \textbf{$p$-value} & \multicolumn{1}{c|}{\textbf{ACC\textsubscript{\textit{m}}}} & \textbf{$p$-value} & \multicolumn{1}{c|}{\textbf{ACC\textsubscript{\textit{m}}}} & \textbf{$p$-value} & \multicolumn{1}{c|}{\textbf{ACC\textsubscript{\textit{m}}}} & \textbf{$p$-value} & \multicolumn{1}{c|}{\textbf{ACC\textsubscript{\textit{m}}}} & \multicolumn{1}{c}{\textbf{$p$-value}} \\ \hline
\textbf{CIFAR-10}                 & \multicolumn{1}{c|}{89.21\%}       & 1.50e-15           & \multicolumn{1}{c|}{86.41\%}       & 9.78e-44           & \multicolumn{1}{c|}{86.68\%}       & 2.98e-54           & \multicolumn{1}{c|}{85.12\%}       & 3.91e-40           & \multicolumn{1}{c|}{79.08\%}       & 4.77e-42           & \multicolumn{1}{c|}{41.25\%}       & 1.00               & \multicolumn{1}{c|}{36.96\%}       & 1.00                                    \\ \hline
\textbf{Imagenette}               & \multicolumn{1}{c|}{80.89\%}       & 1.56e-45           & \multicolumn{1}{c|}{81.74\%}       & 1.10e-38           & \multicolumn{1}{c|}{80.75\%}       & 6.60e-49           & \multicolumn{1}{c|}{81.74\%}       & 8.96e-38           & \multicolumn{1}{c|}{77.83\%}       & 8.44e-31           & \multicolumn{1}{c|}{55.40\%}       & 1.00               & \multicolumn{1}{c|}{46.18\%}       & 1.00                                    \\ \hline
\end{tabular}}
\end{table*}

\mypara{Robustness against Watermark Removal.} To evaluate the adversarial robustness of \name, we further consider the settings where the attacker attempts to remove the watermark through model fine-tuning and pruning. It is worth noting that the attacker is assumed to have access to the partial original pre-training dataset. We conduct a series of experiments employing the SimCLR-based watermarked encoders.

In the attack scenario permitting full fine-tuning of all layers, the ownership verification results (as shown in TABLE~\ref{tab:eaas_robustness_pruning}) indicate that the $p$-values slightly increase after FT but remain significantly below the predefined threshold of 0.05. This demonstrates the strong robustness of the proposed \name against FT attacks. We further perform pruning attacks on the stolen DT encoder with pruning ratios ranging from 5\% to 99\%. As shown in TABLE~\ref{tab:eaas_robustness_pruning}, the ACC\textsubscript{\textit{m}} of the \name watermarked encoder gradually decreases as the pruning ratio increases. When the pruning ratio is below 40\%, the $p$-values remain nearly identical to those of the unpruned model. At PR-60\%, the $p$-values increase slightly but still stay well below the 0.05 threshold. Only when the pruning ratio reaches 85\%, does the ownership verification of \name fail, indicating that the attacker has successfully removed the watermark. However, at this point, the encoder's utility severely degrades, and its accuracy drops to 41.25\% on CIFAR-10 and 55.40\% on Imagenette, rendering the encoder useless and thereby defeating the attacker's goal of maintaining usability.

\begin{table*}[htbp]
\centering
\caption{Comparison of \name and Dziedzic \textit{et al.}~\cite{dziedzic2022difficulty} with robustness against watermark removal.}
\label{tab:eaas_sota} 
\scalebox{0.8}{
\begin{tabular}{c|cc|cc|cc|c|cc}
\hline
\multirow{2}{*}{\begin{tabular}[c]{@{}c@{}}\textbf{Pre-training}\\ \textbf{Dataset}\end{tabular}} & \multicolumn{2}{c|}{\name} & \multicolumn{2}{c|}{Dziedzic \textit{et al.}~\cite{dziedzic2022difficulty}} & \multicolumn{2}{c|}{\textbf{PR-70\%}} & \multirow{2}{*}{\begin{tabular}[c]{@{}c@{}}\textbf{Fine-tuning}\\ \textbf{Dataset}\end{tabular}} & \multicolumn{2}{c}{\textbf{FT}} \\ \cline{2-7} \cline{9-10} 
& \multicolumn{1}{c|}{ACC$_m$$\downarrow$} & $p$-value & \multicolumn{1}{c|}{ACC$_m$$\downarrow$} & $p$-value & \multicolumn{1}{c|}{$p$-value(\name)} & $p$-value(\cite{dziedzic2022difficulty}) & & \multicolumn{1}{c|}{$p$-value(\name)} & $p$-value(\cite{dziedzic2022difficulty}) \\ \hline
\multirow{3}{*}{\textbf{CIFAR-10}} & \multicolumn{1}{c|}{\multirow{3}{*}{\textcolor{blue}{1.3\%}}} & \multirow{3}{*}{3.43e-26} & \multicolumn{1}{c|}{\multirow{3}{*}{\textcolor{red}{14.89\%}}} & \multirow{3}{*}{1.67e-25} & \multicolumn{1}{c|}{\multirow{3}{*}{\textcolor{blue}{3.23e-15}}} & \multirow{3}{*}{\textcolor{red}{1.00}} & \textbf{CIFAR-10} & \multicolumn{1}{c|}{\textcolor{blue}{1.90e-15}} & \textcolor{red}{1.00} \\ \cline{8-10} 
& \multicolumn{1}{c|}{} & & \multicolumn{1}{c|}{} & & \multicolumn{1}{c|}{} & & \textbf{CINIC} & \multicolumn{1}{c|}{\textcolor{blue}{2.38e-15}} & \textcolor{red}{1.00} \\ \cline{8-10} 
& \multicolumn{1}{c|}{} & & \multicolumn{1}{c|}{} & & \multicolumn{1}{c|}{} & & \textbf{STL-10} & \multicolumn{1}{c|}{\textcolor{blue}{8.72e-13}} & \textcolor{red}{1.00} \\ \hline
\end{tabular}}
\end{table*}

\par We further compare our \name with the watermarking method in~\cite{dziedzic2022difficulty} using SimCLR-pretrained encoders on the CIFAR-10. Our analysis shows three main limitations of~\cite{dziedzic2022difficulty}: First, its ownership verification depends on the encoder’s intermediate representations, which restricts its applicability to EaaS scenarios and hinders its extension to broader MLaaS scenarios. Second, as shown in TABLE~\ref{tab:eaas_sota}, the watermark in~\cite{dziedzic2022difficulty} is fragile to removal via fine-tuning with only a small set of samples (e.g., 1,000 images) and becomes ineffective under a pruning ratio of just 75\%. In contrast, \name demonstrates strong robustness against such watermark-removal attacks. Finally, the watermark introduced in~\cite{dziedzic2022difficulty} leads to a pronounced utility degradation (\text{ACC$_m$}$\downarrow$=14.89\%) compared to \name (\text{ACC$_m$}$\downarrow$=1.3\%). We attribute this to two key factors: (1) the method requires a watermark-embedding dataset on a scale comparable to the encoder’s original training data, and (2) it relies on learning discriminative responses to specific, private augmentations (e.g., rotation)—a requirement that fundamentally conflicts with the objective of self-supervised learning methods like SimCLR, which aim to maintain representation invariance under a wide variety of augmentations.

\mypara{Adaptive Attacks.} Under the EaaS scenario, we follow the experimental settings described in Section~\ref{6} for the MLaaS scenario. First, we evaluate the $p$-value of the \name under overwriting and unlearning attacks launched by the adaptive attackers, respectively. As shown in TABLE~\ref{tab:eaas_robustness_k12}, under both adaptive attacks, \name can still successfully extract the embedded watermarks from the suspect encoders, with all $p$-values remaining far below the threshold of 0.05.

\begin{table}[htbp]
\centering
\caption{Knowledgeable Attackers I (Overwriting) and II (Unlearning) under EaaS scenario.}
\label{tab:eaas_robustness_k12} 
\scalebox{0.85}{
\begin{tabular}{c|cc|cc}
\hline
\multirow{2}{*}{\textbf{Dataset}} & \multicolumn{2}{c|}{\textbf{Overwriting}} & \multicolumn{2}{c}{\textbf{Unlearning}} \\ \cline{2-5} 
                                  & \multicolumn{1}{c|}{\textbf{ACC\textsubscript{\textit{m}}}} & \textbf{$p$-value} & \multicolumn{1}{c|}{\textbf{ACC\textsubscript{\textit{m}}}} & \textbf{$p$-value} \\ \hline
\textbf{CIFAR-10}                 & \multicolumn{1}{c|}{84.79\%} & 1.84e-38 & \multicolumn{1}{c|}{78.25\%} & 8.74e-64 \\ \hline
\textbf{Imagenette}               & \multicolumn{1}{c|}{78.05\%} & 8.70e-21 & \multicolumn{1}{c|}{79.15\%} & 4.80e-29 \\ \hline
\end{tabular}}
\end{table}

\begin{figure}[htbp]
    \centering
    \subfloat[]{\includegraphics[width=0.25\textwidth]{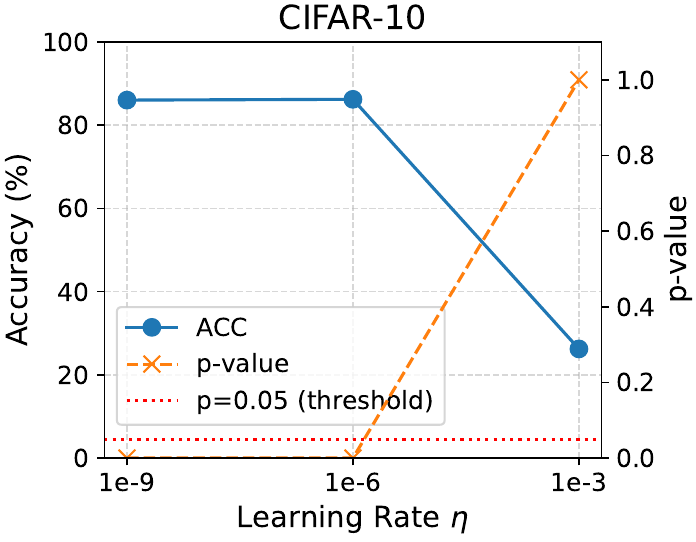}\label{fig:eaas_cifar10}}
    \subfloat[]{\includegraphics[width=0.25\textwidth]{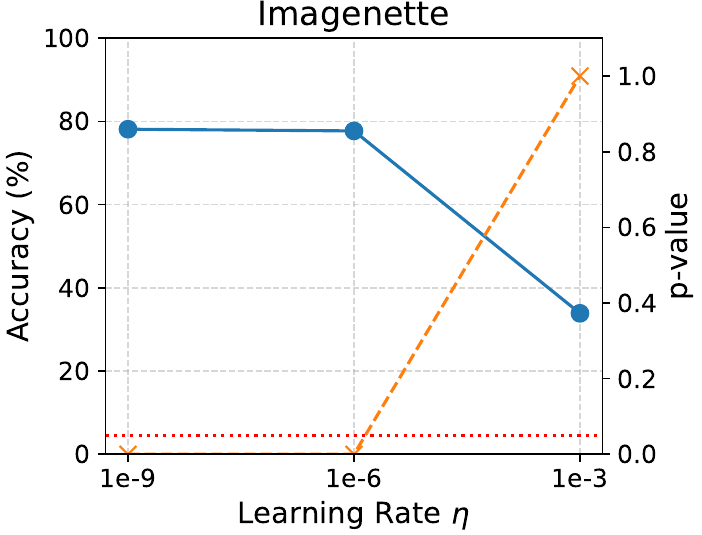}\label{fig:eaas_imagenette}}
    \vspace{0.1in}
    \caption{Knowledgeable Attacker (III) under EaaS scenario.}
    \label{fig:adaptive_eaas_k3}
\end{figure}

Next, unlike in Section~\ref{6}, when simulating the Knowledgeable Attacker (III), we fix $\psi=0.1$ in Eq.~\ref{eq.6} and assess the encoder's utility across different learning rates $\eta$ (i.e., $1 \times 10^{-9}$, $1 \times 10^{-6}$, $1 \times 10^{-3}$) in terms of ACC\textsubscript{\textit{m}} and $p$-value, as illustrated in Fig.~\ref{fig:adaptive_eaas_k3}. The results show that when $\eta=1 \times 10^{-9}$ and $1 \times 10^{-6}$, the ACC\textsubscript{\textit{m}} and $p$-values of the \name encoder on CIFAR-10 and Imagenette remain almost unaffected. However, when $\eta=1 \times 10^{-3}$, the encoder's ACC\textsubscript{\textit{m}} drops sharply to an unusable level, and watermark verification fails simultaneously. These findings indicate that while a large learning rate may destroy the embedded watermark, it also severely impairs the encoder's functionality, thereby defeating the attacker's theft objective.

\subsection{Experimental Results of \name's Ownership Verification Under DINOv2}\label{sec:appendix_dinov2}

\par
\begin{figure}[htbp]
  \centering
    \includegraphics[width=0.4\textwidth]{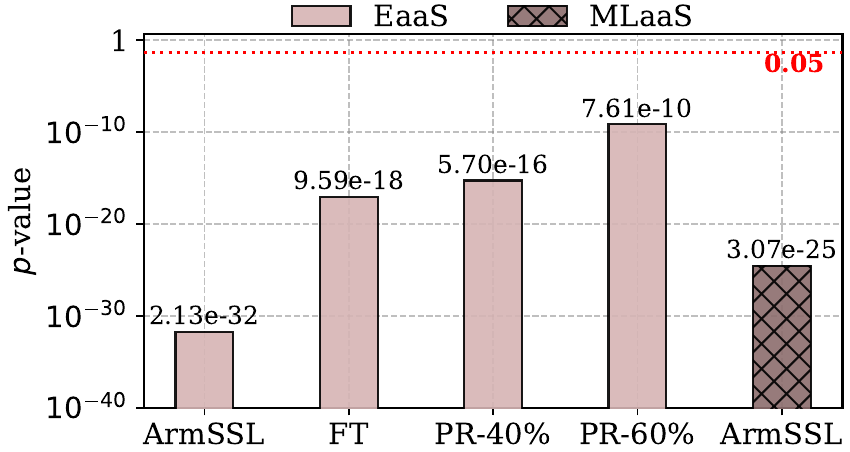}
  \caption{The \name's $p$-value under DINOv2 algorithm.}
  \label{fig:dinov2}
\end{figure}

 To further validate the broad applicability and backbone-agnostic nature of \name, we extend our evaluation to the SOTA SSL framework DINOv2 based on ViT architecture. As illustrated in Fig.~\ref{fig:dinov2}, we evaluate the ownership verification performance ($p$-value) across both EaaS and MLaaS scenarios: (1) Effectiveness: For the watermarked DINOv2 encoder, \name achieves an extremely low $p$-value (i.e., 2.13e-32 in the EaaS scenario and 3.07e-25 in the MLaaS scenario), providing an unambiguous ownership signal that stays significantly below the decision threshold $0.05$; (2) Robustness against fine-tuning: After aggressive full fine-tuning, the embedded watermark remains resilient with a $p$-value of 9.59e-18, confirming that the IP signal is not easily erased; (3) Robustness against model pruning: Under structured pruning, \name continues to exhibit high reliability. At the 40\% and 60\% pruning rate, the corresponding $p$-values remain highly significant. In summary, these results demonstrate that \name is highly effective not only for CNN-based contrastive learning but also for ViT-based self-distillation frameworks, ensuring robust IP protection across various SSL paradigms.

\end{appendices}


\begin{thebibliography}{10}
\providecommand{\url}[1]{#1}
\csname url@samestyle\endcsname
\providecommand{\newblock}{\relax}
\providecommand{\bibinfo}[2]{#2}
\providecommand{\BIBentrySTDinterwordspacing}{\spaceskip=0pt\relax}
\providecommand{\BIBentryALTinterwordstretchfactor}{4}
\providecommand{\BIBentryALTinterwordspacing}{\spaceskip=\fontdimen2\font plus
\BIBentryALTinterwordstretchfactor\fontdimen3\font minus \fontdimen4\font\relax}
\providecommand{\BIBforeignlanguage}[2]{{%
\expandafter\ifx\csname l@#1\endcsname\relax
\typeout{** WARNING: IEEEtran.bst: No hyphenation pattern has been}%
\typeout{** loaded for the language `#1'. Using the pattern for}%
\typeout{** the default language instead.}%
\else
\language=\csname l@#1\endcsname
\fi
#2}}
\providecommand{\BIBdecl}{\relax}
\BIBdecl

\bibitem{chen2020simple}
T.~Chen, S.~Kornblith, M.~Norouzi \emph{et~al.}, ``A simple framework for contrastive learning of visual representations,'' in \emph{Proc. Int. Conf. Mach. Learn. (ICML)}, 2020, pp. 1597--1607.

\bibitem{gui2024survey}
J.~Gui, T.~Chen, J.~Zhang \emph{et~al.}, ``A survey on self-supervised learning: Algorithms, applications, and future trends,'' \emph{IEEE Trans. Pattern Anal. Mach. Intell.}, vol.~46, no.~12, pp. 9052--9071, 2024.

\bibitem{qi2024hssas}
S.~Qi, D.~Wang, Y.~Fan \emph{et~al.}, ``Hssas: Optimizing nlp models by combining self-supervised learning and neural architecture search,'' in \emph{Proc. Int. Conf. Robot. Autom. Intell. Control (ICRAIC)}, 2024, pp. 71--75.

\bibitem{DeepDriving2015}
C.~Chen, A.~Seff, A.~L. Kornhauser \emph{et~al.}, ``Deepdriving: Learning affordance for direct perception in autonomous driving,'' in \emph{Proc. ICCV}.\hskip 1em plus 0.5em minus 0.4em\relax {IEEE} Computer Society, 2015, pp. 2722--2730.

\bibitem{seer_cost}
Y.~LeCun and I.~Misra, ``https://ai.meta.com/blog/self-supervised-learning-the-dark-matter-of-intelligence,'' March 4, 2021.

\bibitem{cong2022sslguard}
T.~Cong, X.~He, and Y.~Zhang, ``Sslguard: A watermarking scheme for self-supervised learning pre-trained encoders,'' in \emph{ACM Conf. Comput. Commun. Secur. (CCS)}, 2022, pp. 579--593.

\bibitem{liu2022stolenencoder}
Y.~Liu, J.~Jia, H.~Liu \emph{et~al.}, ``Stolenencoder: stealing pre-trained encoders in self-supervised learning,'' in \emph{ACM Conf. Comput. Commun. Secur. (CCS)}, 2022, pp. 2115--2128.

\bibitem{dziedzic2022difficulty}
A.~Dziedzic, N.~Dhawan, M.~A. Kaleem \emph{et~al.}, ``On the difficulty of defending self-supervised learning against model extraction,'' in \emph{International Conference on Machine Learning}.\hskip 1em plus 0.5em minus 0.4em\relax PMLR, 2022, pp. 5757--5776.

\bibitem{lv2022ssl}
P.~Lv, P.~Li, S.~Zhu \emph{et~al.}, ``Ssl-wm: A black-box watermarking approach for encoders pre-trained by self-supervised learning,'' \emph{Proc. Netw. Distrib. Syst. Secur. Symp. (NDSS)}, 2024.

\bibitem{wu2022watermarking}
Y.~Wu, H.~Qiu, T.~Zhang \emph{et~al.}, ``Watermarking pre-trained encoders in contrastive learning,'' in \emph{Proc. Int. Conf. Data Intell. Secur. (ICDIS)}.\hskip 1em plus 0.5em minus 0.4em\relax IEEE, 2022, pp. 228--233.

\bibitem{feng2023detecting}
S.~Feng, G.~Tao, S.~Cheng \emph{et~al.}, ``Detecting backdoors in pre-trained encoders,'' in \emph{Proc. IEEE/CVF Conf. Comput. Vis. Pattern Recognit. (CVPR)}, 2023, pp. 16\,352--16\,362.

\bibitem{wang2024mm}
H.~Wang, Z.~Xiang, D.~J. Miller \emph{et~al.}, ``Mm-bd: Post-training detection of backdoor attacks with arbitrary backdoor pattern types using a maximum margin statistic,'' in \emph{Proc. IEEE Symp. Secur. Priv. (SP)}, 2024, pp. 1994--2012.

\bibitem{kolouri2019generalized}
S.~Kolouri, K.~Nadjahi, U.~Simsekli \emph{et~al.}, ``Generalized sliced wasserstein distances,'' \emph{Proc. Adv. Neural Inf. Process. Syst. (NeurIPS)}, vol.~32, 2019.

\bibitem{zhao2024comparison}
Z.~Zhao, L.~Alzubaidi, J.~Zhang \emph{et~al.}, ``A comparison review of transfer learning and self-supervised learning: Definitions, applications, advantages and limitations,'' \emph{Expert Syst. Appl.}, vol. 242, p. 122807, 2024.

\bibitem{chen2020improved}
X.~Chen, H.~Fan, R.~Girshick \emph{et~al.}, ``Improved baselines with momentum contrastive learning,'' \emph{arXiv preprint arXiv:2003.04297}, 2020.

\bibitem{grill2020bootstrap}
J.-B. Grill, F.~Strub, F.~Altch{\'e} \emph{et~al.}, ``Bootstrap your own latent-a new approach to self-supervised learning,'' \emph{Proc. Adv. Neural Inf. Process. Syst. (NeurIPS)}, vol.~33, pp. 21\,271--21\,284, 2020.

\bibitem{chen2021exploring}
X.~Chen and K.~He, ``Exploring simple siamese representation learning,'' in \emph{Proc. IEEE/CVF Conf. Comput. Vis. Pattern Recognit. (CVPR)}, 2021, pp. 15\,750--15\,758.

\bibitem{oquab2023dinov2}
M.~Oquab, T.~Darcet, T.~Moutakanni \emph{et~al.}, ``Dinov2: Learning robust visual features without supervision,'' \emph{Trans. Mach. Learn. Res.}, 2024.

\bibitem{uchida2017embedding}
Y.~Uchida, Y.~Nagai, S.~Sakazawa \emph{et~al.}, ``Embedding watermarks into deep neural networks,'' in \emph{Proc. Int. Conf. Multimedia Retr. (ICMR)}, 2017, pp. 269--277.

\bibitem{chen2019deepmarks}
H.~Chen, B.~D. Rouhani, C.~Fu \emph{et~al.}, ``Deepmarks: A secure fingerprinting framework for digital rights management of deep learning models,'' in \emph{Proc. Int. Conf. Multimedia Retr. (ICMR)}, 2019, pp. 105--113.

\bibitem{liu2021watermarking}
H.~Liu, Z.~Weng, and Y.~Zhu, ``Watermarking deep neural networks with greedy residuals.'' in \emph{Proc. Int. Conf. Mach. Learn. (ICML)}, 2021, pp. 6978--6988.

\bibitem{namba2019robust}
R.~Namba and J.~Sakuma, ``Robust watermarking of neural network with exponential weighting,'' in \emph{ACM Conf. Comput. Commun. Secur. (CCS)}, 2019, pp. 228--240.

\bibitem{pegoraro2024deepeclipse}
A.~Pegoraro, C.~Segna, K.~Kumari, and A.-R. Sadeghi, ``Deepeclipse: How to break white-box dnn-watermarking schemes,'' in \emph{Proc. USENIX Secur. Symp. (USENIX Security ’24)}, 2024, pp. 5287--5304.

\bibitem{jia2021entangled}
H.~Jia, C.~A. Choquette-Choo, V.~Chandrasekaran \emph{et~al.}, ``Entangled watermarks as a defense against model extraction,'' in \emph{30th USENIX security symposium (USENIX Security 21)}, 2021, pp. 1937--1954.

\bibitem{li2025move}
Y.~Li, L.~Zhu, X.~Jia \emph{et~al.}, ``Move: Effective and harmless ownership verification via embedded external features,'' \emph{IEEE Trans. Pattern Anal. Mach. Intell.}, 2025.

\bibitem{shao2024explanation}
S.~Shao, Y.~Li, H.~Yao \emph{et~al.}, ``Explanation as a watermark: Towards harmless and multi-bit model ownership verification via watermarking feature attribution,'' \emph{Proc. Netw. Distrib. Syst. Secur. Symp. (NDSS)}, 2025.

\bibitem{yang2021robust}
P.~Yang, Y.~Lao, and P.~Li, ``Robust watermarking for deep neural networks via bi-level optimization,'' in \emph{Proc. IEEE Int. Conf. Comput. Vis. (ICCV)}, 2021, pp. 14\,841--14\,850.

\bibitem{mehta2022aime}
D.~Mehta, N.~Mondol, F.~Farahmandi \emph{et~al.}, ``Aime: Watermarking ai models by leveraging errors,'' in \emph{Proc. Des. Autom. Test Eur.(DATE)}.\hskip 1em plus 0.5em minus 0.4em\relax IEEE, 2022, pp. 304--309.

\bibitem{xie2025dataset}
Y.~Xie, J.~Song, M.~Xue, H.~Zhang, X.~Wang, B.~Hu, G.~Chen, and M.~Song, ``Dataset ownership verification in contrastive pre-trained models,'' \emph{Proc. Int. Conf. Learn. Represent. (ICLR)}, 2025.

\bibitem{dziedzic2022dataset}
A.~Dziedzic, H.~Duan, M.~A. Kaleem \emph{et~al.}, ``Dataset inference for self-supervised models,'' \emph{Proc. Adv. Neural Inf. Process. Syst. (NeurIPS)}, vol.~35, pp. 12\,058--12\,070, 2022.

\bibitem{dubinski2023bucks}
J.~Dubi{\'n}ski, S.~Pawlak, F.~Boenisch \emph{et~al.}, ``Bucks for buckets (b4b): Active defenses against stealing encoders,'' \emph{Advances in Neural Information Processing Systems}, vol.~36, pp. 55\,237--55\,259, 2023.

\bibitem{shetty2024warden}
A.~Shetty, Y.~Teng, K.~He, and Q.~Xu, ``Warden: Multi-directional backdoor watermarks for embedding-as-a-service copyright protection,'' \emph{Proc. Annu. Meet. Assoc. Comput Linguist.}, pp. 13\,430--13\,444, 2024.

\bibitem{fei2024your}
Z.~Fei, B.~Yi, J.~Geng \emph{et~al.}, ``Your fixed watermark is fragile: Towards semantic-aware watermark for eaas copyright protection,'' \emph{arXiv e-prints}, pp. arXiv--2411, 2024.

\bibitem{tang2023watermarking}
Y.~Tang, J.~Yu, K.~Gai \emph{et~al.}, ``Watermarking vision-language pre-trained models for multi-modal embedding as a service,'' \emph{arXiv preprint arXiv:2311.05863}, 2023.

\bibitem{chen2022effective}
W.~Chen, B.~Wu, and H.~Wang, ``Effective backdoor defense by exploiting sensitivity of poisoned samples,'' \emph{Proc. Adv. Neural Inf. Process. Syst. (NeurIPS)}, pp. 9727--9737, 2022.

\bibitem{cai2013distributions}
T.~Cai, J.~Fan, and T.~Jiang, ``Distributions of angles in random packing on spheres,'' \emph{J. Mach. Learn. Res.}, vol.~14, no.~1, pp. 1837--1864, 2013.

\bibitem{hartigan1979algorithm}
J.~A. Hartigan and M.~A. Wong, ``Algorithm as 136: A k-means clustering algorithm,'' \emph{Journal of the royal statistical society. series c (applied statistics)}, vol.~28, no.~1, pp. 100--108, 1979.

\bibitem{ester1996density}
M.~Ester, H.-P. Kriegel, J.~Sander, X.~Xu \emph{et~al.}, ``A density-based algorithm for discovering clusters in large spatial databases with noise,'' in \emph{kdd}, vol.~96, no.~34, 1996, pp. 226--231.

\bibitem{tao2024distribution}
G.~Tao, Z.~Wang, S.~Feng \emph{et~al.}, ``Distribution preserving backdoor attack in self-supervised learning,'' in \emph{Proc. IEEE Symp. Secur. Priv. (SP)}, 2024, pp. 2029--2047.

\bibitem{larsen2005introduction}
R.~J. Larsen and M.~L. Marx, ``An introduction to mathematical statistics (hoboken, nj,'' 2005.

\bibitem{dai2025division}
Z.~Dai, Y.~Gao, B.~Kuang \emph{et~al.}, ``Division and union: Latent model watermarking,'' \emph{IEEE Transactions on Information Forensics and Security}, 2025.

\bibitem{jia2022badencoder}
J.~Jia, Y.~Liu, and N.~Z. Gong, ``Badencoder: Backdoor attacks to pre-trained encoders in self-supervised learning,'' in \emph{Proc. IEEE Symp. Secur. Priv. (SP)}.\hskip 1em plus 0.5em minus 0.4em\relax IEEE, 2022, pp. 2043--2059.

\bibitem{li2025machine}
N.~Li, C.~Zhou, Y.~Gao \emph{et~al.}, ``Machine unlearning: Taxonomy, metrics, applications, challenges, and prospects,'' \emph{IEEE Trans. Neural Netw. Learn. Syst.}, 2025.

\bibitem{ImageNet}
J.~Deng, W.~Dong \emph{et~al.}, ``Imagenet: A large-scale hierarchical image database,'' in \emph{Proc. CVPR}, 2009, pp. 248--255.

\bibitem{tiny-imagenet}
Y.~Le and X.~Yang, ``Tiny imagenet visual recognition challenge,'' \emph{CS 231N}, vol.~7, no.~7, p.~3, 2015.

\bibitem{imagenette}
``https://tensorflow.google.cn/datasets/catalog/imagenette.''

\bibitem{cifar100}
``https://tensorflow.google.cn/datasets/catalog/cifar100.''

\bibitem{cifar10}
G.~H. A.~Krizhevsky \emph{et~al.}, ``Learning multiple layers of features from tiny images.'' 2009.

\bibitem{coates2011analysis}
A.~Coates, A.~Ng, and H.~Lee, ``An analysis of single-layer networks in unsupervised feature learning,'' in \emph{Proceedings of the fourteenth international conference on artificial intelligence and statistics}.\hskip 1em plus 0.5em minus 0.4em\relax JMLR Workshop and Conference Proceedings, 2011, pp. 215--223.

\bibitem{darlow2018cinic}
L.~N. Darlow, E.~J. Crowley, A.~Antoniou, and A.~J. Storkey, ``Cinic-10 is not imagenet or cifar-10,'' \emph{arXiv preprint arXiv:1810.03505}, 2018.

\bibitem{svhn}
``http://ufldl.stanford.edu/housenumbers.''

\bibitem{stallkamp2011german}
J.~Stallkamp, M.~Schlipsing \emph{et~al.}, ``The german traffic sign recognition benchmark: a multi-class classification competition,'' in \emph{The 2011 international joint conference on neural networks}.\hskip 1em plus 0.5em minus 0.4em\relax IEEE, 2011, pp. 1453--1460.

\end{thebibliography}
\end{document}